\documentclass[acmsmall,screen]{acmart}
\usepackage{multirow}
\usepackage{graphicx}
\usepackage{diagbox}
\usepackage{subcaption}
\usepackage{xspace}
\usepackage[ruled, vlined, linesnumbered]{algorithm2e}
\usepackage{bbding}
\usepackage{threeparttable}
\usepackage{amsmath}
\usepackage{adjustbox}
\usepackage{listings}
\usepackage[T1]{fontenc}
\usepackage{semantic}
\usepackage{newfloat}
\usepackage{booktabs}

\usepackage{makecell}

\usepackage{pifont}

\newcommand{\CodeIn}[1]{{\small\texttt{#1}}}

\DeclareFloatingEnvironment[fileext=frm,placement={!ht},name=Rules]{infrule}

\usepackage[noend]{algpseudocode}
\usepackage[normalem]{ulem}
\useunder{\uline}{\ul}{}
\AtBeginDocument{%
  \providecommand\BibTeX{{%
    \normalfont B\kern-0.5em{\scshape i\kern-0.25em b}\kern-0.8em\TeX}}}
    
\newcommand\detector{VulLibMiner}
\newcommand\javadata{VulLib}

\newcommand\verajava{VeraJava}

\usepackage{xcolor}

\begin{document}

\title{Identifying Vulnerable Third-Party Java Libraries from Textual Descriptions of Vulnerabilities and Libraries}

\author{Tianyu Chen}
\affiliation{%
  \institution{Key Lab of HCST (PKU), MOE; SCS; Peking University}
  \country{China}}
\email{tychen811@pku.edu.cn}

\author{Lin Li}
\email{lilin88@huawei.com}
\affiliation{%
  \institution{Huawei Cloud Computing Technologies Co., Ltd.}
  \city{Beijing}
  \country{China}
}

\author{Bingjie Shan}
\email{shanbingjie@huawei.com}
\affiliation{%
  \institution{Huawei Cloud Computing Technologies Co., Ltd.}
  \city{Beijing}
  \country{China}
}

\author{Guangtai Liang}
\email{liangguangtai@huawei.com}
\affiliation{%
  \institution{Huawei Cloud Computing Technologies Co., Ltd.}
  \city{Beijing}
  \country{China}
}

\author{Ding Li}
\email{ding_li@pku.edu.cn}
\affiliation{%
  \institution{Key Lab of HCST (PKU), MOE; SCS; Peking University}
  \city{Beijing}
  \country{China}
}

\author{Qianxiang Wang}
\email{wangqianxiang@huawei.com}
\affiliation{%
  \institution{Huawei Cloud Computing Technologies Co., Ltd.}
  \city{Beijing}
  \country{China}
}

\author{Tao Xie}
\email{taoxie@pku.edu.cn}
\affiliation{%
  \institution{Key Lab of HCST (PKU), MOE; SCS; Peking University}
  \city{Beijing}
  \country{China}
}

\renewcommand{\shortauthors}{Chen, et al.}

\begin{abstract}
To address security vulnerabilities arising from third-party libraries, security researchers maintain databases monitoring and curating vulnerability reports.
Application developers can identify vulnerable libraries by directly querying the databases with their used libraries.
However, the querying results of vulnerable libraries are not reliable due to the incompleteness of vulnerability reports.
Thus, current approaches model the task of identifying vulnerable libraries as a named-entity-recognition (NER) task or an extreme multi-label learning (XML) task.
These approaches suffer from highly inaccurate results in identifying vulnerable libraries with complex and similar names, e.g., Java libraries.
To address these limitations, in this paper, we propose \detector{}, the first to identify vulnerable libraries from textual descriptions of both vulnerabilities and libraries, together with \javadata{}, a Java vulnerability dataset with their affected libraries.
\detector{} consists of a TF-IDF matcher to efficiently screen out a small set of candidate libraries and a BERT-FNN model to identify vulnerable libraries from these candidates effectively.
We evaluate \detector{} using four state-of-the-art/practice approaches of identifying vulnerable libraries on both their dataset named \verajava{} and our \javadata{} dataset.
Our evaluation results show that \detector{} can effectively identify vulnerable libraries with an average F1 score of 0.657 while the state-of-the-art/practice approaches achieve only 0.521. 

\end{abstract}



 \maketitle

\section{Introduction}~\label{sec:intro}

In software development, third-party libraries play a crucial role and are widely used~\cite{wang2020empirical}. However, their usage also adds to the responsibility of the application developers to address security vulnerabilities arising from the libraries used~\cite{kasauli2021requirements, chen2020machine, meng2018secure, pham2010detection}. To illustrate, developers must update any vulnerable libraries to their latest versions~\cite{prana2021out} and contact the library developers to fix the vulnerabilities. 
Security researchers maintain databases that monitor and curate vulnerability data to assist application developers in identifying vulnerable libraries. 
An example of such a database is the National Vulnerability Database (NVD)~\cite{nvd}. A report in the NVD outlining a vulnerability includes its identification number in the Common Vulnerability Enumeration (CVE) entry, a description of the vulnerability, and the related libraries in the Common Platform Enumeration (CPE) entry~\cite{cpe}. 
Developers can identify vulnerable libraries by directly querying each description and CPE entry with the name of the libraries used. However, the current manual practice of building the NVD is labor-intensive and error-prone. Recent studies have shown that 53.3\% of the vulnerability reports do not mention any affected libraries~\cite{lightxml}, and 59.82\% of the included vulnerable libraries are incomplete and incorrect~\cite{viem}. Therefore, an automated approach to identify libraries affected by a CVE is desirable.


To automatically identify vulnerable libraries, the key idea of existing approaches is to utilize vulnerability descriptions and library names, and these approaches fall into two categories, named-entity-recognition (NER)-based approaches~\cite{viem, anwar2021cleaning, jo2022vulcan, kuehn2021ovana} and extreme multi-label learning (XML)-based approaches~\cite{fastxml, lightxml, chronos}.
NER-based approaches identify affected libraries by matching (or searching) entities (tokens) extracted from vulnerability descriptions to the names of library packages. 
XML-based approaches use machine-learning classifiers to learn the mapping from a given vulnerability's description to a set of vulnerable library names. 
With the help of training on labeled datasets, these XML-based approaches can identify affected libraries more accurately than NER-based approaches.

Unfortunately, utilizing vulnerability descriptions and only library names are vulnerable to complex and similar library names, which makes existing approaches inaccurate in identifying vulnerable libraries.
Take Listing~\ref{lst: maven-corpus} as an example, library \CodeIn{org.jenkins-ci.plugins:job-direct-mail} has seven different tokens after being split by non-alphabetic characters where most of them occurred in other libraries, e.g., \CodeIn{org.jenkins-ci.plugins:mailcomander}. 
With NER-based methods, they fail to distinguish these libraries with complex and similar names.
For example, they might identify other libraries as affected if tokens in their names are also included in the vulnerability description.
Therefore, they cannot distinguish such libraries even if they achieved a high accuracy (e.g., an F1 score of 0.9) in extracting entities from vulnerability descriptions.
Meanwhile, XML-based methods also confuse libraries with similar names.
Distinguishing these libraries with similar names requires XML-based approaches to train on a large number of vulnerabilities.
However, the number of vulnerabilities is relatively smaller than the number of libraries, thus leading to the low accuracy of XML-based approaches.
For example, there are more than 300,000 libraries in Maven~\cite{maven} and more than 400,000 libraries in Pypi~\cite{pypi} while only 7,665 vulnerabilities in the dataset used by existing work~\cite{fastxml, lightxml, chronos}.


Considering only library names is particularly problematic for Java libraries, whose names are more complex and similar than those of other programming languages.
For complexity, our empirical study shows that the average length and token number of Java library names are three times of other programming languages, e.g., Python or JavaScript.
As for similarity, the names of Java libraries consist of group IDs and artifact IDs, so libraries with the same groups or the same artifacts could be more difficult to distinguish.
For example, 1,247 libraries have the same group IDs, \CodeIn{org.jenkins-ci.plugins}.

In this paper, we propose a novel approach to address the limitations of relying solely on package names to identify vulnerable libraries. We suggest considering the descriptions of libraries, as they contain more detailed information. Understanding the context of library descriptions can avoid confusion caused by similar package names. By matching the descriptions of libraries to CVE descriptions, we can develop a method that can handle new libraries not in the training dataset. Based on our insight, we introduce \detector{}, a novel vulnerable library identifier that models the problem as an entity-linking~\cite{shen2014entity} task. It measures the similarity between the CVE description and each library description. 

The main challenge in identifying vulnerable libraries based on their descriptions is to balance efficiency and effectiveness.
On the one hand, we want a large language model that accurately comprehends the semantics of the descriptions. 
On the other hand, identifying the affected libraries of one vulnerability with the help of library descriptions is time-consuming due to invoking the large language model many times (the number of libraries).
For example, there are more than 300,000 Java libraries in Maven, and correlating each of them with one given vulnerability is a huge burden on both efficiency and computation resources.


Our solution to this challenge is to use a lightweight and efficient technique to effectively exclude libraries (e.g., 512 ones from all 311,233 ones) that are apparently unrelated, even though it may not accurately identify all libraries affected by a CVE. 
We propose this solution based on our observation that vulnerability descriptions and library descriptions share the same entities that indicate the group IDs and artifact IDs.
For example, in Listing~\ref{lst: cve-2020-2318} and~\ref{lst: maven-corpus}, both the vulnerability and affected library mentions \CodeIn{jenkins}, \CodeIn{mail}, and \CodeIn{command}.
Thus, libraries that share no/few common entities with the description of a CVE are not affected by this vulnerability.
Building on this insight, we propose a two-stage approach. 
In the first stage, we use a weighted TF-IDF technique to screen candidates of libraries that may be affected by a CVE where the affected libraries are rarely removed in this step. 
In the second stage, we employ a BERT-FNN model, a precise model with relatively high costs, to precisely identify the vulnerable libraries.


In addition to the preceding challenge, our approach also faces the issue of inadequate training data. The current datasets available do not contain any library descriptions, making them unsuitable for training the language models used in \detector{}. To overcome this challenge, we create \verajava{} by extracting Java vulnerabilities and incorporating library descriptions into the existing dataset~\cite{fastxml, lightxml, chronos}.
Considering the limited number (only 948 vulnerabilities with 985 affected libraries) of Java vulnerabilities in \verajava{}, we construct a larger dataset, \javadata{}, with manual validation.
The \javadata{} dataset is created by thoroughly examining the Github Advisory~\cite{githubAD} and NVD~\cite{nvd} database.
Overall, we have gathered 948 Java vulnerabilities that correspond to 925 libraries in \verajava{} and 2,789 vulnerabilities that correspond to 2,095 Java libraries in \javadata{}.

We evaluate \detector{} using four state-of-the-art/practice approaches of vulnerable library identification (FastXML~\cite{fastxml}, LightXML~\cite{lightxml}, Chronos~\cite{chronos}, and our TF-IDF matcher) on both \verajava{} and \javadata{} dataset.
Our evaluation results show a number of findings to demonstrate our \detector{}'s effectiveness and efficiency.
\detector{} effectively achieves the F1 score of 0.657 in identifying vulnerable Java libraries while the state-of-the-art/practice approaches achieves only 0.521.
Additionally, \detector{} is generalizable in identifying zero-shot libraries, which do not occur in the training set.
\detector{} substantially increases the F1 score in identifying zero-shot libraries by 29.5\%.
\detector{} is highly efficient, identifying one vulnerability in less than two seconds.

In summary, this paper makes the following main contributions:
\begin{itemize}
\item We propose to consider descriptions of libraries to accurately identify vulnerable libraries.
\item We propose a novel two-staged approach that can efficiently identify vulnerable libraries while maintaining accuracy. 
\item We construct an open-source dataset, \javadata{}, manually validated by senior software engineers, including 5,028 $\langle CVE, Library \rangle$ pairs.  
\item We conduct a comprehensive evaluation of demonstrating \detector{}'s effectiveness and efficiency, achieving the average F1 score of 0.657 while the state-of-the-art/practice approaches achieve only 0.521.
\end{itemize}

\section{Background}\label{sec:background}
The objective of identifying vulnerable libraries is to locate the libraries that are affected by a particular CVE description. An example of this is provided in Listing~\ref{lst: cve-2020-2318}.
Our goal is to extract the affected library's name, such as \CodeIn{org.jenkins-ci.plugins:mailcommander}.
This process involves analyzing the information provided in the \textit{Description} field.


\subsection{The Limitation of Existing approaches}
Current approaches that identify libraries affected by vulnerabilities can be classified into two categories, NER-based approaches and XML-based approaches.
However, both these approaches rely on only library names for identification, thus plagued in identifying vulnerable libraries with complex and similar names, such as Java libraries.
For example, \CodeIn{org.jenkins-ci.plugins:mailcommander} and \CodeIn{org.jenkins-ci.plugins:mailer} are two different libraries that are not affected by the same vulnerabilities while these two categories of approaches cannot distinguish them correctly.
We summarize these two categories of techniques and explain their limitations in the following sub-sections.

\subsubsection{NER-Based Approaches}~\label{sec: ner approaches}
Named-entity-recognition (NER)-based approaches~\cite{viem, anwar2021cleaning, jo2022vulcan, kuehn2021ovana} collect keywords from a vulnerability description and match them to library names of libraries. 
One example is VIEM~\cite{viem}, which uses a Bi-GRU model~\cite{mikolov2013distributed} for NER.
In Listing~\ref{lst: cve-2020-2318}, tokens extracted include "Jenkins-ci," "Mail," "Commander," "Plugin," "job," and "controller." 
Then these keywords are used to match library names. 
For example, \CodeIn{org.jenkins-ci.plugins:job-direct-mail} is considered vulnerable because it matches the keywords "plugin," "job," and "mail."

It can be difficult to accurately identify Java libraries using NER-based approaches because the library names of two Java libraries can be very similar.
In Listing~\ref{lst: maven-corpus}, we show that these approaches classify Libraries (1-4) as vulnerable because most tokens in their names belong to library entities extracted by the NER model. However, only Library (2) is vulnerable out of the four.  NER-based methods mistakenly classify Libraries (1), (2), and (4) as vulnerable because they contain keywords extracted from the CVE description, such as "mail", "job", and "jenkins-ci".

\subsubsection{XML-Based Approaches}
Existing extreme multi-label learning (XML) based approaches~\cite{fastxml, lightxml, chronos} take the vulnerable library identification problem as a multi-label classification problem. Assuming there are $K$ libraries in the market, XML-based approaches take the description of a CVE as input and classify it into $K$ categories that correspond to $K$ libraries. 

When compared to NER-based approaches, approaches that use labeled datasets based on similar vulnerability descriptions in the training set achieve higher accuracy~\cite{fastxml, lightxml}. 
However, XML-based approaches are not effective in identifying libraries with similar names due to a lack of sufficient training vulnerabilities.
For example, there are more than 300,000 libraries in Maven~\cite{maven} and more than 400,000 libraries in Pypi~\cite{pypi} while only 7,665 vulnerabilities in the dataset used by existing work~\cite{fastxml, lightxml, chronos}.
Thus, XML-based approaches cannot distinguish these libraries with similar names.
For instance, in Listing~\ref{lst: maven-corpus}, the XML-based approaches incorrectly identify \CodeIn{org.jenkins-ci.plugins:mailer} as the affected library for CVE-2020-2318, while the actual answer is \CodeIn{org.jenkins-ci.plugins:mailcommander}.
The reason for this incorrect classification is that \CodeIn{org.jenkins-ci.plugins:mailer} is seen as the label of another vulnerability (CVE-2018-8718) during the training of their classification models while \CodeIn{org.jenkins-ci.plugins:mailcommander} is an unseen new label.




\subsubsection{Our Insight}
We have observed that NER-based and XML-based methods fail to consider library descriptions, which contain important details for matching CVE descriptions to affected libraries. Our proposal in this paper is to include library descriptions to enhance the precision of identifying vulnerable Java libraries. We use the example in Listing~\ref{lst: maven-corpus} to explain our insight.
The names of Library (1), (2), and (3) are quite similar as they have the same group ID and similar artifact IDs.
However, their descriptions are different, thus enabling us to identify the correct vulnerable library.
The description of Library (1) mentions \CodeIn{``parent POM project''}, which is not relevant to the vulnerability's input description. The description of Library (3) repeats its name, which does not provide any additional correlation to the vulnerability's input description. However, the description of Library (2) mentions \CodeIn{``read a mail''}, which is similar in semantics to \CodeIn{``viewed by users''}. Therefore, Library (2) is more likely to be identified as vulnerable when compared to Library (1) and (3).

Based on the above observation, we convert the vulnerable library identification problem to an entity-linking problem, which correlates a vulnerability description with a library description.
Given a pair of $\langle CVE, Library \rangle$, we propose an approach that detects whether the library description matches the CVE description. If so, the library is potentially affected by the CVE. 

\begin{figure}[t]
\begin{lstlisting}[caption= CVE-2020-2318,label=lst: cve-2020-2318, language=java]
CVE: CVE-2020-2318,
Description: "Jenkins Mail Commander Plugin for Jenkins-ci Plugin 1.0.0 and earlier stores passwords unencrypted in job config.xml files on the Jenkins controller where they can be viewed by users with Extended Read permission, or access to the Jenkins controller file system."

Hyperlink: "https://www.jenkins.io/security/advisory/2020-11-04/#SECURITY-2085"

CWE: [CWE-522: "Insufficiently Protected Credentials"
      CWE-256: "Plaintext Storage of a Password"]

Labels (CPE): ["org.jenkins-ci.plugins:mailcommander"]
\end{lstlisting}
\end{figure}

\begin{figure}[t]
\begin{lstlisting}[caption= Vulnerable libraries of CVE-2020-2318, label=lst: maven-corpus, language=java]
Candidates of existing NER-based approaches: 
    (1) "org.jenkins-ci.plugins:mailer",
    (2) "org.jenkins-ci.plugins:mailcommander"
    (3) "org.jenkins-ci.plugins:job-direct-mail"
    ......
Output of Existing NER-based approaches: [(1), (2), (3)]
---------------------------------------------------------------------------------
Candidates of existing XML-based approaches: 
    (1) "org.jenkins-ci.plugins:mailer",
    (2) "org.jenkins-ci.plugins:mailcommander"
    (3) "org.jenkins-ci.plugins:job-direct-mail"
    ......
Output of Existing XML-based approaches: [(1)]
---------------------------------------------------------------------------------
//Group IDs are omitted here.
Candidates with library descriptions: 
    (1) mailer: "The Jenkins Plugins Parent POM Project"
    (2) mailcommander: "This plug-in provides function that read a mail subject as a CLI Command"
    (3) job-direct-mail: "Job Direct Mail Plugin"
    ......
Output with library descriptions: [(2)]
\end{lstlisting}
\end{figure}

\subsection{Challenges of Identifying Vulnerable Libraries with Descriptions}
We face a challenge in balancing efficiency and effectiveness in our approach.
To fully understand the vulnerability and library descriptions, we require a large language model, as shallow and small models are not as effective in comprehending natural language~\cite{bert}. 
However, we also need a fast algorithm to process the vast number of CVEs and libraries. 
Without the first step of filtering, we need to invoke the effectiveness model 311,233 times to identify the affected libraries of one vulnerability.
If we use the popular BERT model, which takes about 4 ms to process a $\langle CVE, Library \rangle$, it would take around 20 minutes for each vulnerability.
Such an expensive time cost is unacceptable to respond to a user query of identifying a given vulnerability~\cite{zhang2023bidirectional}.
Additionally, considering the rapid increase of vulnerabilities (66 vulnerabilities each day on average~\cite{nvd2022report}), it might cost 22 hours to identify all their affected libraries, which is also a substantial cost.

\subsubsection{The high-level idea of addressing the preceding dilemma}
Our proposed approach involves a two-staged process for identifying vulnerable libraries. 
Specifically, we efficiently screen out a small set of candidate libraries that contain the affected libraries before using an effective model to precisely identify the vulnerable libraries, thus reducing the total time consumption.
This is based on our observation that vulnerability descriptions and library descriptions share the same entities that indicate the group IDs and artifact IDs.
For group IDs, the entities of group IDs have few ambiguities because they directly indicate their library belongings.
Even the vulnerability reports do not include the complete group IDs, e.g., \CodeIn{io.jenkins.plugins}, it mentions keywords that indicate its developers, e.g., \CodeIn{jenkins} or \CodeIn{apache}.
As for artifact IDs, they work as the module names imported by users.
When writing vulnerability descriptions, reporters tend to include the vulnerable modules directly.

This two-staged approach significantly reduces the processing time of the effective model.
Even with a large candidate size of 512, we can still reduce the processing time of one vulnerability from 20 minutes to 2 seconds approximately.
Thus, the time consumption of identifying the affected libraries of all new libraries might decrease from 22 hours to 3 minutes, which is acceptable for a vulnerability database.

\section{Approach}\label{sec:approach}
In this section, we propose \detector{}, the first approach to identify vulnerable libraries from descriptions of both vulnerabilities and libraries.
Unlike conventional NER-based and XML-based approaches, our approach can correlate the natural language descriptions of libraries to the description of vulnerabilities, introducing higher identification accuracy.

\begin{figure*}[t]
\centering
\includegraphics[width=1\linewidth]{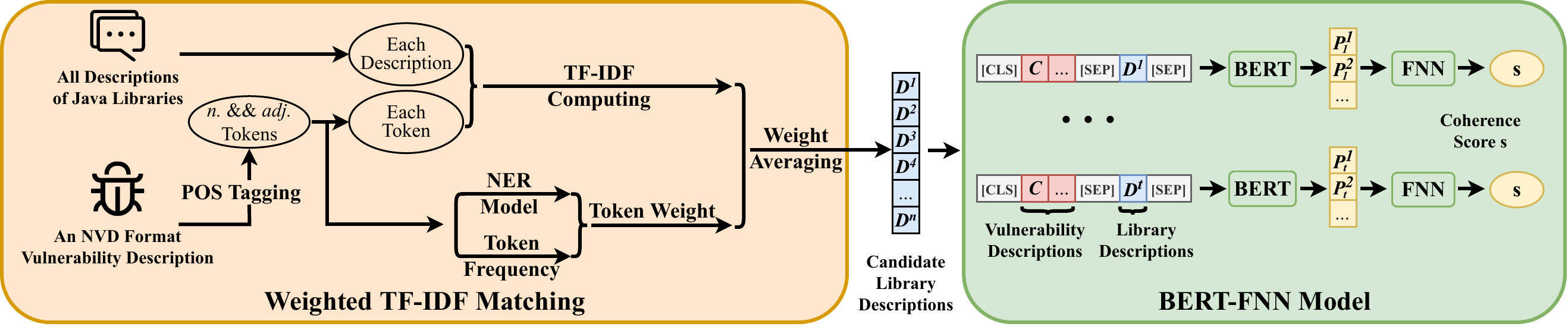}
\caption{High-level framework of \detector{}}
\label{fig: framework}
\end{figure*}

As depicted in Figure~\ref{fig: framework}, \detector{} takes the textual descriptions of one vulnerability and all libraries as inputs and outputs each library's possibility of being affected.
To ensure effectiveness, \detector{} adopts a widely-used language model, BERT-FNN, which generates the probability of each library being affected by a given vulnerability. 
To further improve efficiency, before running the BERT-FNN model, \detector{} first scans the libraries with a TF-IDF matcher.
This step removes the libraries that are obviously unrelated to a given vulnerability with few false negatives, thus substantially reducing the number of libraries that require further analysis by the BERT-FNN model.

\subsection{Weighted TF-IDF Matching}
To efficiently screen potentially vulnerable libraries with their descriptions, we design our weighted TF-IDF matching algorithm to reduce the number of libraries to be tested by the large language models.
This algorithm screens out a small set of candidate libraries that have a higher probability of being affected by a given vulnerability.
The underlying principle is that a library affected by a vulnerability will exhibit a higher degree of textual similarity in its description to that of this vulnerability.
Thus, to efficiently identify this similarity, we calculate the weighted TF-IDF score between the description of a vulnerability and a library. 
On a high level, the weighted TF-IDF scores measure the frequency and significance of tokens by the description of a library and a vulnerability.

Our technique is shown in Algorithm~\ref{alg: TF IDF}.
It takes the description of a specific vulnerability (in the form of NVD) and all the descriptions of Java libraries as inputs.
Here, we concatenate library descriptions with their names (in short, library descriptions in the following of this paper).
Then, it outputs \CodeIn{CandidateNum} libraries as the set of candidate libraries, which contains the affected libraries with high probability.
This algorithm screens out candidate libraries through two steps: preprocessing, and weighted TF-IDF averaging.

\subsubsection{Preprocessing}
In Lines 1-7 of Algorithm~\ref{alg: TF IDF}, we preprocess the input vulnerability description with two steps: Part-of-speech tagging (POS tagging)~\cite{schmid1994part}, and weight computing.
The input of preprocessing is a vulnerability description in the form of a token sequence.
The output of POS tagging is a set of noun and adjective tokens extracted from the input description, and the output of weight computing each token's weight for averaging TF-IDF scores.

In Lines 1-2, we use the third-party POS tagging algorithm in the python library, \CodeIn{nltk.tag} with a set of [NN, NNS, NNP, NNPS, JJ, JJR, JJS], corresponding to noun and adjective tokens.
The rationale behind this step is that noun, and adjective tokens in the description of a vulnerability are likely to be correlated with a vulnerable library or its functionalities.
In this step, we do not introduce any domain knowledge of vulnerabilities and libraries, and we will take advantage of them in the next step.

In Lines 3-7, we compute each token's weight screened out in the preceding step.
We define each token's weight as its frequency in the given vulnerability description.
We amplify its weight if one token belongs to a library-name entity as they are highly indicative of the affected libraries.
These entities are extracted by a BERT-based NER model modified from a third-party one~\cite{finkel2005incorporating}.
We train this NER model to determine whether the tokens in a vulnerability description belong to a library-name entity.
We use 2,196 vulnerability descriptions for training and 548 vulnerability descriptions for testing. 
Our targeted entities are derived by BERT's~\cite{bert} default tokenizer applied to all library names in Maven~\cite{maven}.
The precision, recall, and F1 score of the NER model is 96.16\%, 92.85\%, and 94.48\%, respectively.
Then in Lines 5-7, we compute the frequency of each token and amplify the weights of library-name entities with a constant factor \CodeIn{EntityWeight}.
Based on our evaluation results in Section~\ref{sec: hyper-parameter}, \CodeIn{EntityWeight}$= 4$ is a proper value.

\subsubsection{Weighted TF-IDF averaging}
The term frequency/inverse document frequency (TF-IDF)~\cite{jones1972statistical, jones2004IDF} is a commonly used technique to rank documents (e.g., library descriptions) by weighting a given term (token) in each document.
TF-IDF is formally defined as follows:
\begin{equation}~\label{eq: TF IDF}
    TF\mathcal{-}IDF[i, j] = TF[i, j] \times \log ({\frac{N}{DF[i] + 1}})
\end{equation}
Here $TF\mathcal{-}IDF[i, j]$ is the weight of term $i$ in a document $j$, $N$ is the number of documents in the collection, $TF[i, j]$ is the term frequency of term $i$ in document $j$, and $DF[i]$ is the document frequency of term $i$ in the collection.

In Lines 8-13 of Algorithm~\ref{alg: TF IDF}, we calculate the weighted TF-IDF scores for each token in the vulnerability description against each library description.
Specifically, we first compute the TF value of term $i$ and document $j$, and the IDF value of term $i$, respectively in Lines 11 and 12.
Then, we compute the average weight of TF-IDF scores based on the weights assigned to each token during preprocessing.
This weighted averaging approach ensures that the results are not skewed by the length of library descriptions. 
Specifically, the TF value is normalized by the description length, while the IDF value remains constant for each description.



\begin{algorithm}[t]
	\SetKwData{DiffUtils}{DiffUtils}\SetKwData{integer}{\textbf{int}}
	\SetKwFunction{Union}{Union}\SetKwFunction{getDepth}{minDepth}
	\SetKwInOut{Input}{Input}\SetKwInOut{Output}{Output}
 
	\Input{ $vulnDes$, a given vulnerability description.}
        \Input{ $libDes$, a list of library descriptions.}
	\Output{$candidates$, the indexes of TopK candidate libraries}
	\BlankLine

        \tcp{Preprocessing}
        $tags \gets \mbox{[NN, NNS, NNP, NNPS, JJ, JJR, JJS]}$ \tcp*[h]{nouns and adjectives } \;
        $vulnTokens \gets POStag(vulnDes, tags)$ \;
        $entites \gets BERT\mathit{-}NER(vulnTokens)$\;
        \For{$term \in vulnTokens$}{
            $weight[i] \gets Frequency(vulnDes, term)$\;
            \If{$entites.contains(term)$}{
                $weight[i] \gets \mbox{\CodeIn{EntityWeight}} \times weight[i]$;
            }
        }
        
        \tcp{Weighted TF-IDF Averaging}
        \For{$j \in (0, libDesc.size())$}{
            \For{$i \in (0, vulnTokens.size())$}{
                $term \gets vulnTokens.getToken(i)$\;
                $TF[i, j] \gets \frac{Frequency(libDes[j], term)}{libDes[j].size()}$\;
                $IDF[i] \gets \log \frac{libDes.size()}{\sum_{j} libDes[j].contains(term)}$
            }
            $score[j] \gets \sum_{i} \frac{weight[i]}{\sum_{i} weight[i]} \times TF[i, j] \times IDF[i]$\;
        }
        $candidates \gets IndexOf(score.Top(\mbox{\CodeIn{CandidateNum}}))$\;
        \Return{$candidates$}\;
\caption{Weighted TF-IDF Matching}\label{alg: TF IDF}
\end{algorithm}

\subsection{BERT-FNN Model}
This section details the design and functionality of our BERT-FNN model, crucial for precisely identifying affected libraries of a vulnerability.
In this step, we leverage the pre-trained language model, BERT~\cite{bert}, to comprehend and correlate the descriptions of vulnerabilities and libraries.
The positive samples for training are the affected libraries among the candidate libraries while the rest of them are trained as negative samples.
Additionally, BERT is a lightweight model that does not depend on a large number of training data, so 238,080 ($1860 \times 128$) pairs of $\langle CVE, Library \rangle$ are sufficient for model training.

The structure of this model is illustrated in Figure~\ref{fig: framework}.
This model takes one pair of descriptions <vulnerability description, library description> as inputs and outputs a coherence score.
This score represents the likelihood of the library being affected by this vulnerability.
The model comprises a BERT encoder and a feed-forward neural network (FNN) layer.

Taking the t-th candidate library as an example, our BERT encoder concatenates the vulnerability description, $C$, the library description, $D^{t}$, and BERT's special placeholders, \CodeIn{[CLS]} and \CodeIn{[SEP]}, as inputs:
\begin{equation}
    input(C, D^{t}) = \mbox{\CodeIn{[CLS]} } C \mbox{ \CodeIn{[SEP]} } D^{t} \mbox{ \CodeIn{[SEP]}}
\end{equation}


The output of BERT is a series of embedding results of the input descriptions, $[P_{1}^{t}, P_{2}^{t}, \dots P_{n}^{t}]$:
\begin{equation}
    [P_{1}^{t}, P_{2}^{t}, \dots P_{n}^{t}] = BERT(input(C, D^{t}))
\end{equation}
where $n$ denotes the length of embedding results.

The FNN layer utilizes the Sigmoid function~\cite{sigmoid} to calculate the coherence score $s^t$ from the preceding embedding results.
\begin{equation}
    \begin{aligned}
        & \hat{s}^{t} = FNN([P_{1}^{t}, P_{2}^{t}, \dots P_{n}^{t}]) \\
        & s^{t} = \frac{1}{1 + exp(\hat{s}^{t}) }
    \end{aligned}
\end{equation}

The model employs weighted binary cross-entropy~\cite{de2005tutorial} as its loss function.
For the input vulnerability, it takes the coherence scores of all candidate libraries, $\{s^t\}$, and their labels, $\{y^t\}$, as inputs: 
\begin{equation}~\label{eq: loss}
    Loss(s, y) = -\frac{1}{T} \sum_{t = 1}^{T} ( \alpha * y^{t} log (s^{t}) + (1 - \alpha) * (1 - y^{t}) log (1 - s^{t}) )
\end{equation}

In Equation~\ref{eq: loss}, $y^{t} = 1$ signifies that the t-th library is affected by the input vulnerability, and otherwise $y^{t} = 0$.
Additionally, we amplify the weight of positive samples (correct <CVE, Library> pairs) with a factor $\alpha = 0.9$ to alleviate the imbalance of training data (positive samples count for less than 1\% in training data).
\section{Dataset}~\label{sec: dataset}
One of the practical challenges in realizing our approach is acquiring adequate training data. 
Although previous approaches~\cite{fastxml, lightxml, chronos} have released their datasets, none of them can be used in our approach because they do not contain descriptions of libraries. 
Besides, the existing datasets do not comprehensively cover known vulnerabilities of open-source Java libraries. 

To address these limitations, we develop two specific datasets.
The first, \verajava{}, extends the dataset from previous work~\cite{fastxml, lightxml, chronos} and contains 1,899 $\langle CVE, Library \rangle$ pairs. 
Considering that the original dataset does not contain library descriptions, we manually add the descriptions to the dataset. 
The second dataset, \javadata{}, is a comprehensive dataset built from GitHub Advisory~\cite{githubAD} and the National Vulnerability Database (NVD)~\cite{nvd}.
\javadata{} contains all confirmed $\langle CVE, Library \rangle$ pairs in GitHub Advisory up to March 2022, along with 772 $\langle CVE, Library \rangle$ pairs whose affected libraries are specified as Java libraries in NVD.
Both two datasets are available on our website~\footnote{https://github.com/q5438722/VulLibMiner}.




\subsection{\javadata{} Construction}
\begin{figure}[t]
\centering
\includegraphics[width=\linewidth]{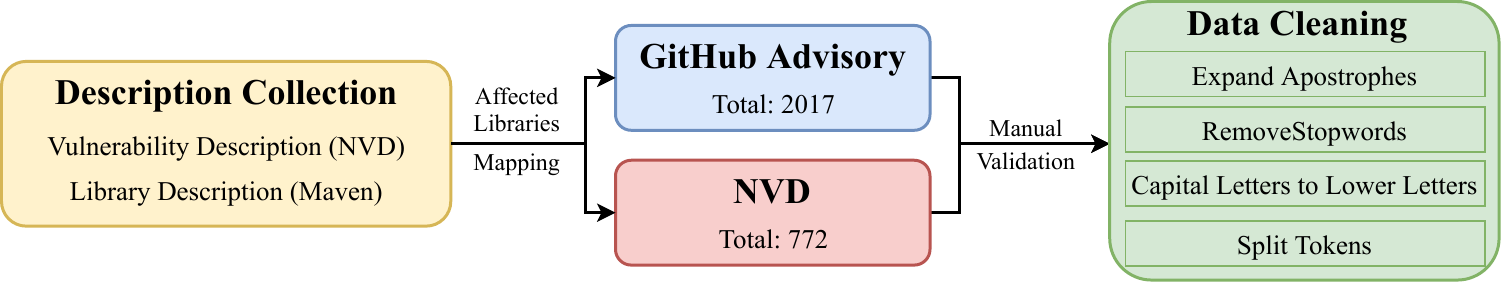}
\caption{Construction Framework of \javadata{}{}}
\label{fig : dataset}
\end{figure}

Figure~\ref{fig : dataset} depicts the three steps of constructing \javadata{}: 
(1) We collect the descriptions of Java vulnerabilities and libraries;
(2) We collect the mapping between vulnerabilities and their affected libraries as positive samples;
(3) We conduct data cleaning over the descriptions of both vulnerabilities and libraries.

\subsubsection{Description collection}~\label{sec: description collection} 
Now that \detector{} takes the descriptions of vulnerabilities and libraries as inputs, we collect these descriptions first.
For vulnerabilities, we choose NVD~\cite{nvd} as the source database. 
We collect all the vulnerability descriptions until 2022.03.31 from NVD.
Here, the latest CVE of vulnerabilities is CVE-2022-27216.
As for libraries, we collect the data of all 435,642 Java libraries from maven~\cite{maven}, and 311,233 of them have descriptions.

\subsubsection{Vulnerability Collection}
After collecting the descriptions, the next step is collecting the affected libraries of vulnerabilities, i.e., collecting $\langle CVE, Library \rangle$ pairs.
We utilize GitHub Advisory~\cite{githubAD} and NVD~\cite{nvd} as our data sources.
We collect the labels of 2,017 Java vulnerabilities from GitHub Advisory and the labels of 772 vulnerabilities from NVD.
Each of these labels is manually confirmed by one engineer.

Specifically, our engineers confirm each pair of $\langle CVE, Library \rangle$ based on two criteria.
First, they check whether the library has reported this vulnerability as an issue or has a code commit fixing this vulnerability.
Second, our engineers check whether the reference links of this vulnerability report refer to this library directly or indirectly.
If one of the preceding criteria is satisfied, our engineers conclude that this vulnerability affects this library.
Additionally, our engineers check the description of each vulnerability to see whether it mentions other libraries that are not included and check these libraries based on the preceding criteria.


\subsubsection{Data cleaning}
To keep the same settings with existing work~\cite{fastxml, lightxml, chronos}, we conduct the following four data-cleaning steps:
\begin{enumerate}
    \item Expand apostrophes in descriptions. 
    \item Change capital letters to lower letters.
    \item Split descriptions by non-alphanumeric characters.
    \item Remove stopwords~\cite{ghag2015comparative}, which are frequently occurring words that hardly carry any information and orientation, such as ``the'', ``is'', ``have''. They are commonly removed in natural language processing tasks~\cite{ghag2015comparative, kaur2018systematic}.
\end{enumerate}

\subsection{Dataset Partition}
For \verajava{}, we follow the same partition as the latest baseline~\cite{chronos}, which includes zero-shot (unseen) libraries in its testing set.
To keep the same settings with our baselines~\cite{fastxml, lightxml, chronos}, we directly use its testing set as its validation set.
As for \javadata{}, we follow the same partition ratio (3:1:1) recorded in these baseline approaches.

\subsubsection{The background information of our manual efforts}
The reliability of \javadata{} highly depends on the expertise of our engineers, so we explain the background of our manual efforts here.
\javadata{} is constructed by 10 software engineers, each with at least two years in Java development, and one year in vulnerability-related tasks, e.g., vulnerability mining and localization.
The entire process of manually verifying vulnerabilities and their affected libraries in \javadata{} takes approximately 50 man-days.


\subsection{Comparison between \javadata{} and \verajava{}}~\label{sec: dataset comparison}

\begin{table}[t]
\centering
\caption{The Distribution of \verajava{}, and \javadata{}}
\label{tab: dataset distribution}
\begin{tabular}{lrrrrrrr}
\toprule
Dataset     & \#<V, L> & \#V & \#L & AVG(L/V) & \multicolumn{1}{c}{Training} & \multicolumn{1}{c}{Validation} & \multicolumn{1}{c}{Testing} \\
\midrule
\verajava{} & 1,899     & 948             & 985       & 2.00   & 629 & 319 & 319   \\
\javadata{} & 5,028     & 2,789            & 2,095      & 1.80   & 1,668 & 556 & 565  \\
\bottomrule
\end{tabular}
 \begin{itemize}
    \item \quad V represents vulnerabilities; L represents libraries.
    \item \quad AVG(L/V) represents the average number of affected libraries per vulnerability.
    \item \quad In \verajava{}, we use its testing set for validation.
 \end{itemize}
\end{table}

Table~\ref{tab: dataset distribution} shows the distribution of \verajava{} and \javadata{}.
One notable observation is that \javadata{} includes more than twice as many vulnerabilities as \verajava{}.
\javadata{} contains 5,028 pairs of $\langle CVE, Library \rangle$,  while \verajava{} contains only 1,899 pairs of $\langle CVE, Library \rangle$.
This disparity indicates that \javadata{} is a more comprehensive dataset for identifying vulnerable libraries.

Another notable observation is that the average number of libraries per vulnerability in \verajava{} and \javadata{} is 2.00 and 1.80.
These statistics indicate that there are a large number of unseen (zero-shot) Java libraries (affected by only one vulnerability).
Identifying zero-shot libraries is inherently challenging due to their absence during model training.
We define the zero-shot percentage as the percentage of vulnerabilities whose affected libraries are all zero-shot libraries.
The zero-shot percentage of \javadata{} and \verajava{} are $28.31\%$ and $47.34\%$, respectively.
These percentages underscore the importance of considering zero-shot libraries in identifying vulnerable Java libraries.

\section{Evaluation}\label{sec:evalution}
Our evaluation answers the following four research questions about \detector{}:
\begin{itemize}
    \item \textbf{RQ1}: How effectively can \detector{} identify Java libraries when compared with baseline approaches? 
    \item \textbf{RQ2}: What is the contribution of identifying zero/full-shot libraries to the achieved effectiveness of \detector{}? 
    \item \textbf{RQ3}: How efficiently can \detector{} identify vulnerable Java libraries?
    \item \textbf{RQ4}: What are the proper values of the hyper-parameters in the TF-IDF matcher?
\end{itemize}


\begin{table}[t]
\centering
\caption{Hyper-parameters used for Training BERT-FNN}
\label{tab: hyper parameter}
\begin{tabular}{llll}
\toprule
Hyper-parameters          & Value & Hyper-parameters          & Value \\
\midrule
Optimizer                 & Adam & Learning Rate & 2e-5      \\
Soft Token Embedding Size & 768  & Weight Decay  & 0.01        \\
Max Sequence Length       & 512  & Batch Size    & 32         \\
\bottomrule
\end{tabular}
\end{table}

\subsection{Baselines}
We select four state-of-the-art/practice (SOTA) baselines that identify the affected libraries from vulnerability descriptions.
We select three XML-based approaches, FastXML~\cite{fastxml}, LightXML~\cite{lightxml}, and Chronos~\cite{chronos}.
We select our TF-IDF matcher, which also includes an effective BERT-based NER model with an F1 score of 94.48\%, as another baseline because VIEM~\cite{viem} can not be reproduced~\footnote{We have contacted them by email and have not received their response. Their repository also has the same issue while not addressed.}
In a recent study~\cite{chronos}, these baselines are more effective than other approaches, such as Bonsai~\cite{khandagale2020bonsai} and ExtremeText~\cite{wydmuch2018no}.

\subsection{Evaluation Environments}
We perform all the evaluations in the environment running on the system of Ubuntu 18.04.
We use one Intel(R) Xeon(R) Gold 6248R@3.00GHz CPU, which contains 64 cores and 512GB memory.
We use one Tesla A100 PCIe GPU with 40GB memory for model training and inference.
The hyper-parameters used for BERT-FNN are listed in Table~\ref{tab: hyper parameter}.

\subsection{Metrics}
We evaluate the performance of library identification models through precision (P), recall (R), and F1-score (F1) upon the Top k ($k=1,2,3$) prediction results.
These metrics are the same as our baselines~\cite{fastxml, lightxml, chronos}, thus ensuring a fair evaluation.
Additionally, they are widely used to evaluate similar multi-label prediction tasks~\cite{wu2023plms, narayan2021discriminative}, such as version identification~\cite{Libid, Atvhunter}.  
For a given vulnerability $v$, a prediction model outputs each library's probability of being affected by $v$.
Given the libraries with Topk probabilities, $prediction_k(v)$, and our ground-truth labels of $v$, $affected(v)$, the precision, recall, and F1 metrics are defined as:

\begin{equation}~\label{eq: precision recall}
    \begin{aligned}
        &precision@k(v) = \frac{|prediction_k(v) \cap affected(v)|}{min(k, affected(v)}\\
        &recall@k(v) = \frac{|prediction_k(v) \cap affected(v)|}{|affected(v)|}\\
        &F1@k(v) = \frac{2 \times precision@k(v) \times recall@k(v)}{precision@k(v) + recall@k(v)}
    \end{aligned}
\end{equation}
The precision, recall, and F1 of \detector{} and our baselines are defined as the arithmetic mean of their values on each vulnerability.


\subsection{\textbf{RQ1}: How effectively can \detector{} identify Java libraries when compared with baseline approaches?}


\subsubsection{Methodology}
We evaluate the effectiveness of \detector{} on both datasets from the perspective of the preceding three metrics, $precision@k$, $recall@k$, and $F1@k$.
Specifically, we use two evaluation scenarios for comparison.
To keep the same setting of baselines, we use the first scenario that \detector{} identifies vulnerable libraries from only affected libraries, i.e., the 948/2,095 libraries affected by the vulnerabilities in \verajava{} or \javadata{}.
However, this setting is limited because it assumes that other Java libraries, which count for more than 99\% among all Java libraries, will not be affected by future vulnerabilities.
To address this limitation, we design the second evaluation scenario that \detector{} identifies vulnerable libraries from all maven libraries (311,233 ones).



\subsubsection{Identification results from only affected libraries}
Table~\ref{tab: top precision recall f1} shows the precision, recall, and F1 score of \detector{} when identifying affected libraries from only affected libraries. 
Here, \detector{} demonstrates high effectiveness among all three metrics in both datasets.
Specifically, \detector{} achieves an average F1 score of 0.724 in \verajava{} and 0.706 in \javadata{}.
In general, this result indicates that \detector{} can effectively identify vulnerable libraries.

When compared with our baselines, \detector{} achieves higher F1@1 scores in both datasets. 
The average F1 scores of \detector{} on \verajava{} and \javadata{} are 0.724 and 0.706 while the best of the baselines' average F1 scores are 0.653 and 0.638, respectively.
The main difference between \detector{} and XML-based approaches is that we consider the description of libraries while FastXML, LightXML, and Chronos only use their names as labels.
As for NER-based approaches, these approaches also struggle due to a lack of correlating the semantics of vulnerabilities and libraries, thus leading to lower F1 scores of 0.490 and 0.493.
On the contrary, \detector{} learns the semantic correlation between vulnerability and library descriptions.
Thus, the improvement in F1 scores indicates that \detector{} is more effective than these baseline approaches.

\begin{table*}[t]
\small
\centering
\caption{Topk Precision, Recall, and F1 Scores from Only Affected Java Libraries}
\label{tab: top precision recall f1}
\begin{tabular}{clcccccccccc}
\toprule
 \multirow{2}{*}{Dataset}     & \multicolumn{1}{c}{\multirow{2}{*}{Approach}} & \multicolumn{3}{c}{Top1} & \multicolumn{3}{c}{Top2} & \multicolumn{3}{c}{Top3} & Avg.\\
\cmidrule(lr){3-5}\cmidrule(lr){6-8}\cmidrule(lr){9-11}\cmidrule(lr){12-12}
                                                           & \multicolumn{1}{c}{}                          & Prec.  & Rec.   & F1     & Prec.  & Rec.   & F1     & Prec.  & Rec.   & F1  &  F1  \\
\midrule
\multirow{5}{*}{VeraJava} & FastXML     & 0.257 & 0.157 & 0.195 & 0.249 & 0.213 & 0.23  & 0.261 & 0.246 & 0.253 & 0.226 \\
                          & LightXML    & 0.320 & 0.198 & 0.245 & 0.276 & 0.243 & 0.259 & 0.289 & 0.278 & 0.283 & 0.262 \\
                          & Chronos     & 0.686 & 0.460 & 0.551 & 0.704 & 0.636 & 0.668 & 0.755 & 0.727 & 0.741 & 0.653 \\
                          & TF-IDF (NER)   & 0.477 & 0.330 & 0.390 & 0.538 & 0.481 & 0.508 & 0.579 & 0.566 & 0.573 & 0.490 \\
\cmidrule(lr){2-12}
                          & VulLibMiner & 0.754 & 0.523 & \textbf{0.618} & 0.785 & 0.703 & \textbf{0.742} & 0.823 & 0.801 & \textbf{0.812} & \textbf{0.724} \\
\midrule
\multirow{5}{*}{VulLib}   & FastXML     & 0.292 & 0.194 & 0.233 & 0.273 & 0.238 & 0.254 & 0.270 & 0.258 & 0.264 & 0.250 \\
                          & LightXML    & 0.450 & 0.327 & 0.378 & 0.450 & 0.409 & 0.428 & 0.468 & 0.452 & 0.460 & 0.422 \\
                          & Chronos     & 0.618 & 0.470 & 0.534 & 0.673 & 0.627 & 0.649 & 0.741 & 0.722 & 0.731 & 0.638 \\
                          & TF-IDF (NER)   & 0.471 & 0.365 & 0.411 & 0.519 & 0.482 & 0.500 & 0.576 & 0.560 & 0.568 & 0.493 \\
\cmidrule(lr){2-12}
                          & VulLibMiner & 0.715 & 0.556 & \textbf{0.626} & 0.747 & 0.696 & \textbf{0.720} & 0.782 & 0.760 & \textbf{0.771} & \textbf{0.706} \\
\bottomrule
\end{tabular}
\end{table*}

\subsubsection{Identification results from all maven libraries}
Table~\ref{tab: maven precision recall f1} shows the precision, recall, and F1 score of \detector{} when identifying affected libraries from all maven libraries. 
We show that \detector{} also achieves substantially high performance on all three metrics in both datasets. 
\detector{} achieves an average F1 score of 0.621 in \verajava{} and 0.657 in \javadata{}, respectively.
These scores are notably higher than the baseline best F1 scores of 0.443 and 0.521
This superiority indicates that \detector{} can effectively identify vulnerable libraries in a realistic, and wide-ranging library environment.

Additionally, as shown in Table~\ref{tab: top precision recall f1} and Table~\ref{tab: maven precision recall f1}, FastXML and LightXML achieves the same scores in both scenarios.
This consistency is attributed to their inherent limitation that they can identify only libraries encountered during training, leading to their inability to identify zero-shot (unseen) libraries~\cite{chronos}. 

\begin{table*}[t]
\small
\centering
\caption{Topk Precision, Recall, and F1 Scores from All Maven Libraries}
\label{tab: maven precision recall f1}
\begin{tabular}{clcccccccccc}
\toprule
 \multirow{2}{*}{Dataset}     & \multicolumn{1}{c}{\multirow{2}{*}{Approach}} & \multicolumn{3}{c}{Top1} & \multicolumn{3}{c}{Top2} & \multicolumn{3}{c}{Top3} & Avg. \\
\cmidrule(lr){3-5}\cmidrule(lr){6-8}\cmidrule(lr){9-11}\cmidrule(lr){12-12}
                                                           & \multicolumn{1}{c}{}                          & Prec.  & Rec.   & F1     & Prec.  & Rec.   & F1     & Prec.  & Rec.   & F1 
 & F1     \\
\midrule
\multirow{5}{*}{VeraJava} & FastXML     & 0.257 & 0.157 & 0.195 & 0.249 & 0.213 & 0.230 & 0.261 & 0.246 & 0.253 & 0.226\\
                          & LightXML    & 0.320 & 0.198 & 0.245 & 0.276 & 0.243 & 0.259 & 0.289 & 0.278 & 0.283 & 0.262\\
                          & Chronos     & 0.497 & 0.336 & 0.401 & 0.467 & 0.428 & 0.446 & 0.490 & 0.475 & 0.482 & 0.443\\
                          & TF-IDF (NER)   & 0.153 & 0.068 & 0.094 & 0.208 & 0.169 & 0.186 & 0.251 & 0.241 & 0.246 & 0.175\\
\cmidrule(lr){2-12}
                          & VulLibMiner & 0.677 & 0.464 & \textbf{0.551} & 0.661 & 0.588 & \textbf{0.622} & 0.700 & 0.678 & \textbf{0.689} & \textbf{0.621} \\
\midrule
\multirow{5}{*}{VulLib}   & FastXML     & 0.292 & 0.194 & 0.233 & 0.273 & 0.238 & 0.254 & 0.270 & 0.258 & 0.264 & 0.250\\
                          & LightXML    & 0.450 & 0.327 & 0.378 & 0.450 & 0.409 & 0.428 & 0.468 & 0.452 & 0.460 & 0.422\\
                          & Chronos     & 0.516 & 0.400 & 0.451 & 0.547 & 0.514 & 0.530 & 0.588 & 0.576 & 0.582 & 0.521\\
                          & TF-IDF (NER)   & 0.189 & 0.136 & 0.158 & 0.229 & 0.212 & 0.220 & 0.266 & 0.260 & 0.263 & 0.214\\
\cmidrule(lr){2-12}
                          & VulLibMiner & 0.669 & 0.520 & \textbf{0.585} & 0.695 & 0.647 & \textbf{0.670} & 0.724 & 0.705 & \textbf{0.715} & \textbf{0.657} \\
\bottomrule
\end{tabular}
\end{table*}



\subsubsection{A case study of false positives/negatives}
To further understand the scope of \detector{}'s effectiveness, we conduct a case study of\detector{}'s false positives/negatives.
In Listing~\ref{lst: false positive negative}, we show an example, CVE-2019-1003041~\cite{security1353}. 
This vulnerability allows attackers to execute arbitrary scripts and affect two Java libraries, \CodeIn{org.jenkins-ci.plugins:groovy} and \CodeIn{org.jenkins-ci.plugins:script- security}.

\detector{} incorrectly identifies two libraries as vulnerable due to their descriptions' similarity to the vulnerability's description.
For example, the description of \CodeIn{org.jenkins-ci.plugins:pipeline- dependency-walker} mentions ``execute a pipeline task for this job and all its downstream jobs'', which is quite similar to the vulnerability description.
This case study suggests that including more detailed information (descriptions) about this vulnerability and libraries can further reduce these false positives.

As for the false negative library, \CodeIn{org.jenkins-ci.plugins:script-security} is incorrectly excluded by our TF-IDF matcher.
The main reason is that the description of this library shares only two tokens with the given vulnerability description, \CodeIn{jenkins}, and \CodeIn{scripts}, and both of them are common across library names and descriptions.
Specifically, there are 2,083 libraries whose names have token \CodeIn{jenkins} and 1,192 libraries whose names have token \CodeIn{scripts}.
Thus, these two tokens are assigned with low TF-IDF scores, leading to the library's exclusion during TF-IDF matching.

Given the relatively low number of false positives/negatives and considering that \CodeIn{org.jenkins- ci.plugins:script-security} is also neglected by NVD~\cite{nvd} and only identified by \CodeIn{Jenkins} itself, these cases of false positives/negatives are deemed acceptable in real-world applications.



\begin{figure}[t]
\begin{lstlisting}[caption= A Case Study of False Positives/Negatives: CVE-2019-1003041, label=lst: false positive negative, language=java]
CVE: "CVE-2019-1003041",
Description: "A sandbox bypass vulnerability in Jenkins Pipeline: Groovy Plugin 2.64 and earlier allows attackers to invoke arbitrary constructors in sandboxed scripts."
---------------------------------------------------------------------------------
Affected libraries: 
    (1) "org.jenkins-ci.plugins:groovy",
    (2) "org.jenkins-ci.plugins:script-security"
---------------------------------------------------------------------------------
Descriptions of affected libraries: 
    (1) groovy: "Groovy"
    (2) script-security: "Allows Jenkins administrators to control what in-process scripts can be run by less-privileged users"
---------------------------------------------------------------------------------
---------------------------------------------------------------------------------
VulLibMiner`'`s top3 libraries: 
    (1) "org.jenkins-ci.plugins:pipeline-dependency-walker",
    (2) "org.jenkins-ci.plugins:pipeline-maven-parent",
    (3) "org.jenkins-ci.plugins:groovy"
---------------------------------------------------------------------------------
Descriptions of VulLibMiner`'`s top3 libraries: 
    (1) pipeline-dependency-walker: "Plugin allows to execute a pipeline task for the job and all its downstream jobs"
    (2) pipeline-maven-parent: "This plugin provides maven integration with Pipeline by providing a withMaven step"
    (3) groovy: "Groovy"
\end{lstlisting}
\end{figure}




\subsection{\textbf{RQ2}: What is the contribution of identifying zero/full-shot libraries to the achieved effectiveness of \detector{}?}~\label{sec: eq zero}

In this research question, we evaluate \detector{}'s effectiveness in identifying zero-shot and full-shot libraries when compared with baseline approaches.
Here, zero-shot libraries are defined as those that do not occur as the labels of vulnerabilities during training.
Thus, vulnerable libraries can be divided into zero-shot and full-shot ones and zero-shot ones are crucial and challenging to identify.
In Section~\ref{sec: dataset comparison}, we have shown that zero-shot libraries are unavoidable in real-world applications.
They count for about 28.31\% and 47.34\% in \javadata{} and \verajava{}, respectively.
Additionally, in Section~\ref{sec: description collection}, we show that there are 435,642 Java libraries, and all of them might be affected by vulnerabilities in the future.
Additionally, zero-shot libraries do not appear in the training set by definition, thus requiring identification approaches to be generalizable to identify them without any prior knowledge about these libraries.

\subsubsection{Methodology}
We evaluate the effectiveness of \detector{}'s identifying zero/full-shot libraries by dividing the testing set of \javadata{} into two sub-sets: zero-shot and full-shot ones.
A vulnerability is classified as a zero-shot one if all its affected libraries are not present as labels in the training set; otherwise, it is classified as a full-shot one.
Specifically, there are 160 zero-shot vulnerabilities and 405 full-shot ones in \javadata{}.
Considering that a recent study~\cite{chronos} shows that FastXML and LightXML are ineffective in identifying zero-shot libraries, we take Chronos and our TF-IDF matcher for comparison.
We evaluate the effectiveness of \detector{} and baselines on both zero-shot and full-shot libraries of \javadata{}.
Specifically, we also use the preceding three metrics, precision@k, recall@k, and F1@k.



\begin{table*}[t]
\small
\centering
\caption{Topk Precision, Recall, and F1 Scores on Zero-Shot Libraries}
\label{tab: zero full}
\begin{tabular}{clcccccccccc}
\toprule
 \multirow{2}{*}{Dataset}     & \multicolumn{1}{c}{\multirow{2}{*}{Approach}} & \multicolumn{3}{c}{Top1} & \multicolumn{3}{c}{Top2} & \multicolumn{3}{c}{Top3} & Avg. \\
\cmidrule(lr){3-5}\cmidrule(lr){6-8}\cmidrule(lr){9-11}\cmidrule(lr){12-12}
                                                           & \multicolumn{1}{c}{}                          & Prec.  & Rec.   & F1     & Prec.  & Rec.   & F1     & Prec.  & Rec.   & F1 
 & F1     \\
\midrule
\multirow{4}{*}{Zero-Shot}       & Chronos     & 0.352 & 0.317 & 0.334 & 0.485 & 0.475 & 0.480 & 0.529 & 0.525 & 0.527 & 0.447    \\
                            & TF-IDF (NER)   & 0.250 & 0.215 & 0.231 & 0.309 & 0.300 & 0.305 & 0.340 & 0.338 & 0.339 & 0.292 \\
                            & VulLibMiner & 0.544 & 0.485 & 0.512 & 0.594 & 0.581 & 0.587 & 0.641 & 0.635 & 0.638 & 0.579    \\
\cmidrule(lr){2-12}
                            & Improvement & 54.5\%   & 53.0\%   & 53.3\%   & 22.5\%   & 22.3\%   & 22.3\%   & 21.2\%   & 21.0\%   & 21.1\%   & 29.5\%  \\
\midrule
\multirow{4}{*}{Full-Shot}         & Chronos     & 0.597 & 0.448 & 0.512 & 0.598 & 0.554 & 0.575 & 0.624 & 0.609 & 0.616 & 0.568 \\
                            & TF-IDF (NER)   & 0.165 & 0.105 & 0.128 & 0.198 & 0.177 & 0.187 & 0.237 & 0.229 & 0.233 & 0.183 \\
                            & VulLibMiner & 0.719 & 0.535 & 0.613 & 0.735 & 0.673 & 0.702 & 0.758 & 0.733 & 0.745 & 0.687 \\
\cmidrule(lr){2-12}
                            & Improvement & 20.4\%   & 19.4\%   & 19.7\%   & 22.9\%   & 21.5\%   & 22.1\%   & 21.5\%   & 20.4\%   & 20.9\%   & 21.0\%   \\
\bottomrule
\end{tabular}
\end{table*}

\subsubsection{General results}
Table~\ref{tab: zero full} shows the results of zero/full-shot library identification.
The results demonstrate that \detector{} enhances the average F1 score by 21.0\% in zero-shot scenarios and by 29.5\% in full-shot scenarios.
This substantial improvement underscores the effectiveness of \detector{} in identifying both zero-shot and full-shot libraries.


For NER-based approaches, our TF-IDF matcher shows better performance in zero-shot scenarios compared to full-shot ones.
However, its overall F1 scores are still lower than those achieved by \detector{}. 
This disparity is largely attributed to its lack of correlating the semantics of both vulnerability and library descriptions, resulting in lower F1 scores in both scenarios.

As for XML-based approaches, Chronos employs an XML-based zero-shot classifier, ZestXML~\cite{zestxml}, which is designed under the assumption that vulnerabilities with similar descriptions tend to be associated with similarly named libraries.
Although Chronos can identify zero-shot libraries by leveraging correlations from its training set, it still struggles to distinguish libraries with similar names. 
Consequently, its average F1 score of zero-shot library identification is only 0.447 while that of \detector{} is 0.579, highlighting the \detector{}'s effectiveness in identifying zero-shot libraries.

\subsection{\textbf{RQ3}: How efficiently can \detector{} identify vulnerable Java libraries?}~\label{sec: rq efficiency}
This research question evaluates the runtime overhead of \detector{} on \javadata{}.
We measure the end-to-end time of both our TF-IDF matcher and the BERT-FNN model based on the Linux System Call \CodeIn{time}.
For each component, we record the average time of three times' experiments to alleviate the effects of randomness.
According to the results given in Section~\ref{sec: hyper-parameter}, when evaluating the efficiency costs of our TF-IDF matcher, we set its hyper-parameters as follows, \CodeIn{EntityWeight} = 4 and \CodeIn{CandidateNum} = 512.

\subsubsection{General results}
The evaluation result is shown in Table~\ref{tab: time}.
The end-to-end time consumption for \detector{} to identify a given vulnerability is 1.709 seconds on average.
Such a time consumption is acceptable for each query in a software engineering task~\cite{zhang2023bidirectional}.
Additionally, this time consumption is acceptable for vulnerability database maintainers to automatically identify the affected libraries of vulnerabilities.
Recent studies~\cite{nvd2022report} show that there are approximately 24,000 vulnerabilities in 2022, equating to around 66 daily.
This result indicates an approximate daily identification time of 2.8 minutes with \detector{}, a manageable duration even if the number of vulnerabilities increases rapidly in the future. 
Thus, \detector{} is both practical and efficient for identifying vulnerable libraries in real-world scenarios.

In terms of model training, \detector{} is highly efficient.
Our BERT-FNN model costs only 2.5 hours for training and the NER model in our TF-IDF matcher costs only 0.95 hours.
Considering the improvement of effectiveness, \detector{}'s training costs are also acceptable and manageable.


\begin{table}[t]
\centering
\caption{Execution time for training and prediction on \javadata{}}
\label{tab: time}
\begin{tabular}{lrrr}
\toprule
Approach    & Train (h) & Pred. (s) & Avg. Pred. (s) \\
\midrule
\detector{} & 3.45       & 1,082            & 1.709             \\
\midrule
TF-IDF (with NER)      & 0.95         & 78            & 0.123             \\
TF-IDF (without NER)     & -         & 26            & 0.033             \\
BERT-FNN     & 2.50         & 1,004            & 1.586             \\
\bottomrule
\end{tabular}
\end{table}

\subsection{\textbf{RQ4}: What are the proper values of the hyper-parameters in the TF-IDF matcher?}~\label{sec: hyper-parameter}
In this research question, we explore the proper values of two hyper-parameters in our TF-IDF matcher, namely \CodeIn{EntityWeight} and \CodeIn{CandidateNum}.
\CodeIn{EntityWeight} represents the weight of each named entity in the description of each vulnerability when calculating the weight of this vulnerability, and \CodeIn{CandidateNum} refers to the number of candidate libraries screened out by our TF-IDF matcher for further analysis.

\subsubsection{Methodology}
To explore the proper values of these two hyper-parameters, we evaluate the effectiveness of our TF-IDF matcher under various assignments of these two hyper-parameters.
We use $recall@k$ for evaluation.
As defined in Equation~\ref{eq: precision recall}, $recall@k$ indicates the proportion of correctly identified vulnerable libraries out of the total vulnerable libraries when \CodeIn{CandidateNum}$=k$.
The denominator of $recall@K$ is $identify(v)$, the total number of vulnerable libraries, and the numerator of $recall@k$ is $|prediction_k(v) \cap identify(v)|$, the number of identified libraries screened out by the TF-IDF matcher.
Our target is to balance the effectiveness and efficiency cost because both the results of $recall@k$ and the runtime overhead of the BERT-FNN model increase with the size of candidate libraries.
To directly show how these two hyper-parameters influence the end-to-end effectiveness of \detector{}, we also use the average F1 score of \detector{} as another evaluation metric.

\begin{figure}[t]
\begin{minipage}{0.48\textwidth}
    \centering
    \includegraphics[width=1\linewidth]{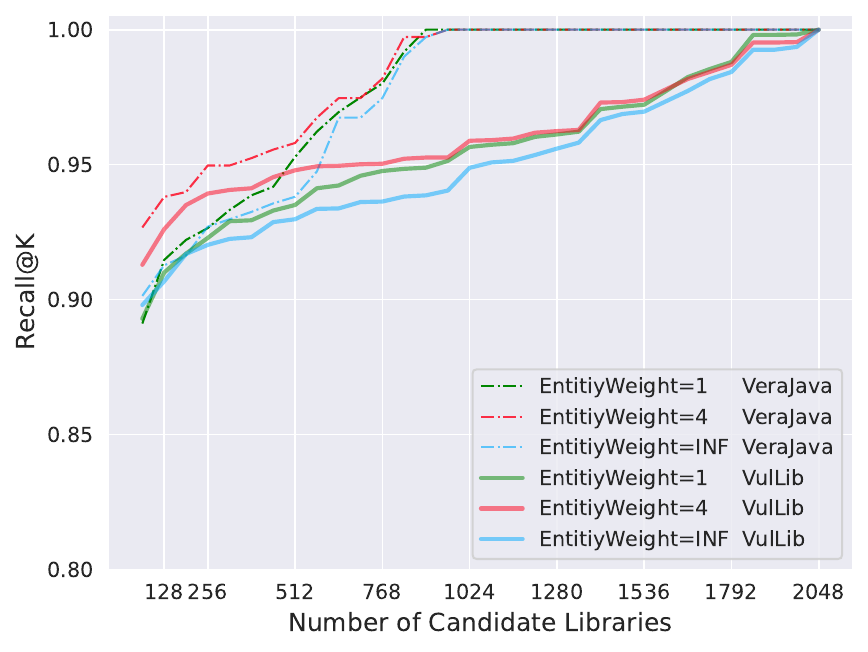}
    \subcaption{From Only Affected Libraries}
    \label{fig: tf-idf top recall}
\end{minipage}
\begin{minipage}{0.48\textwidth}
    \centering
    \includegraphics[width=1.\linewidth]{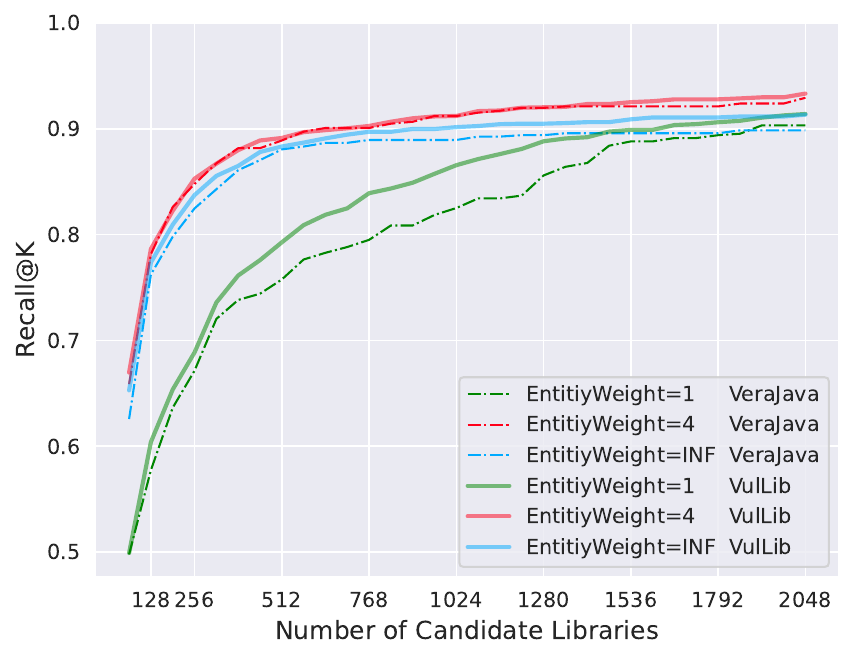}
    \subcaption{From All Maven Libraries}
    \label{fig: tf-idf maven recall}
\end{minipage}
\caption{Recall@k of Our TF-IDF matcher}
\label{fig: tf-idf recall}
\end{figure}

\subsubsection{The proper value of \CodeIn{EntityWeight}}
Figure~\ref{fig: tf-idf top recall} and Figure~\ref{fig: tf-idf maven recall} show the results of $recall@k$ of our TF-IDF matcher.
In both figures and datasets, the red lines (\CodeIn{EntityWeight}$=4$) consistently surpass the green (\CodeIn{EntityWeight}$=1$) and blue (\CodeIn{EntityWeight}$=$\CodeIn{INF}) lines.
Thus, we show that when the \CodeIn{CandidateNum} varies, the \CodeIn{EntityWeight}$=4$ is more effective than $recall@k$ than \CodeIn{EntityWeight}$=1$ and \CodeIn{EntityWeight}$=$\CodeIn{INF} in our TF-IDF matcher.
This improvement mainly comes from our NER model.
The named entities extracted by our NER model are more likely to mention the names and descriptions of libraries, thus helping our TF-IDF matcher identify the affected libraries more effectively.
Meanwhile, a setting of infinity (\CodeIn{INF}) overly prioritizes named entities and neglects other valuable tokens in vulnerability descriptions.
Therefore, \CodeIn{EntityWeight}$=4$ provides an optimal result, effectively leveraging both named entities and other tokens in identifying affected libraries.

\begin{figure}[t]
    \centering
    \includegraphics[width=.8\linewidth]{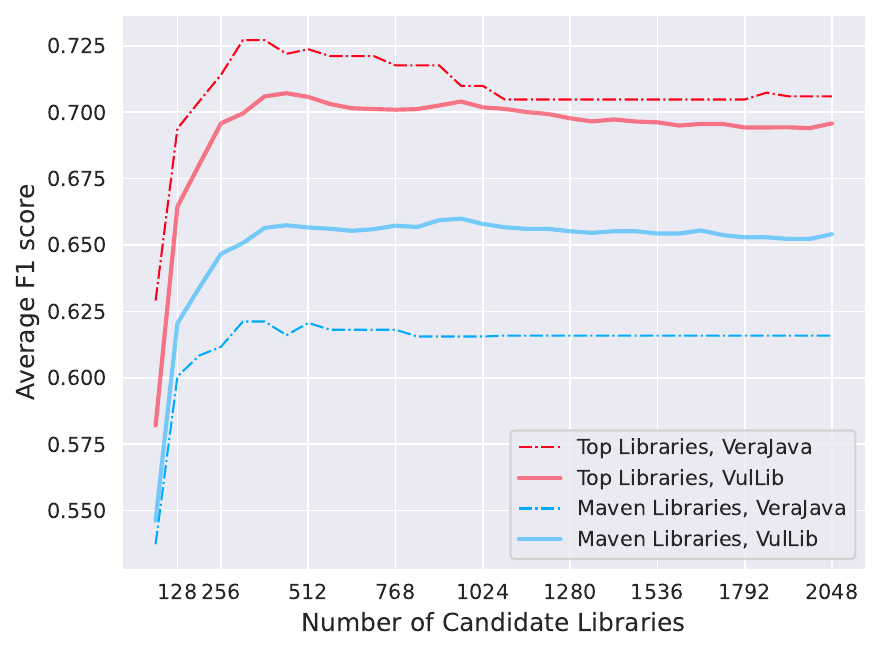}
    \caption{Recall@k of Our TF-IDF matcher}
    \label{fig: tf-idf f1}
\end{figure}

\subsubsection{The proper value of \CodeIn{CandidateNum}}
For the first scenario (identifying libraries from only affected ones) in Figure~\ref{fig: tf-idf maven recall}, we show that the results of $recall@k$ (the red lines) achieve 0.95 when $k$ equals 512 in both datasets.
As for the second scenario (identifying libraries from all maven libraries), the $recall@k$ also achieves at least 0.9 when $k$ equals 512 in both datasets.
Thus, 512 is a proper value for \CodeIn{CandidateNum} as our TF-IDF matcher has less than 10\% false negative libraries.

Figure~\ref{fig: tf-idf f1} also shows the end-to-end impact of varying \CodeIn{CandidateNum} on \detector{}'s overall F1 score.
From this figure, we show that \detector{} achieves the highest end-to-end average F1 score when \CodeIn{CandidateNum} ranges from 256 to 512.
\CodeIn{CandidateNum}$<256$ results in an increasing rate of false negatives, i.e., vulnerable libraries that are not screened out in this step, particularly noticeable when identifying libraries from the entire Maven dataset.
For example, when \CodeIn{CandidateNum}$=128$, the $recall@k$ is only 0.786, which substantially decreases the effectiveness of \detector{}.
Even if our BERT-FNN model is absolutely accurate, its F1 score is still lower than 0.786.
Conversely, \CodeIn{CandidateNum}$>512$ does not significantly enhance \detector{}'s effectiveness and incurs higher computational costs.
This is because our BERT-FNN model needs to process a larger pool of candidate libraries, leading to an increasing runtime and a higher challenge in accurately ranking these candidates. 
Therefore, a \CodeIn{CandidateNum} of 512 is established as a proper setting, in the perspective of \detector{}'s end-to-end effectiveness.

\section{discussion}~\label{sec:discuss}
\subsubsection*{\textbf{Identifying vulnerable libraries of other programming languages}}

\detector{} is designed with a high degree of generalization ability and is not restricted to Java vulnerabilities.
For instance, when applied to Python vulnerabilities, \detector{} can effectively identify affected libraries by training on a dataset of Python $\langle CVE, Library \rangle$ pairs with library descriptions, such as those from PyPI~\cite{pypi}. 
We implement and evaluate \detector{} on Java vulnerabilities due to the complexity and similarity of Java library names, which poses difficulties for current approaches to identify. 
This is in contrast to existing approaches that already demonstrate high F1 scores in identifying vulnerable libraries in languages other than Java~\cite{fastxml, lightxml, chronos}. 
Furthermore, the widespread usage of Java libraries in software development~\cite{wang2020empirical} also underscores the necessity of enhancing vulnerable Java library identification.

 
\subsubsection*{\textbf{Exploiting Large Pre-trained Language Models}}


A natural question is whether Large Pre-trained Language Models (LLMs) can improve \detector{}'s effectiveness.
Although we do believe that LLMs, such as ChatGPT~\cite{chatgpt} or StarCoder~\cite{starcoder}, can perform well in identifying vulnerable libraries, we consider them as future work as they belong to different scopes of approaches.
Additionally, exploiting LLMs leads to a high cost of efficiency and computation resources.
If we use StarCoder instead, the time and computation costs might increase by 50 times as it has 45 times more parameters (15B) than a BERT-based model (340M).
As shown in Section~\ref{sec: hyper-parameter}, \detector{} needs to invoke our BERT-FNN model 512 times while invoking an LLM 512 times is substantially consuming.





\section{Threats to Validity} \label{sec:threats}

A major threat to external validity is the accuracy of vulnerability labels in our \verajava{} and \javadata{} datasets.
For \verajava{}, we follow the same labels as our baseline approaches~\cite{fastxml, lightxml, chronos}.
As for \javadata{}, its labels are collected and verified by GitHub Advisory and NVD maintainers, and then manually verified by our engineers.
Thus, this threat can be minimized.
The threats to internal validity are instrumentation effects that can bias our results. 
The parameters in Table~\ref{tab: hyper parameter} might cause such effects.
To reduce these threats, we follow the same parameters for baseline evaluation and use the default parameters of BERT for \detector{}.
Additionally, the similarity of our baseline evaluation results with those reported in their paper~\cite{chronos} increases the credibility of our evaluation results.
The threat to construct validity mainly comes from the choice of evaluation metrics.
To minimize this threat, we use the same metrics as our baselines to fairly compare \detector{} with these baselines.
Although other metrics, such as Matthews correlation coefficient (MCC)~\cite{yao2020assessing}, and Area Under Curve (AUC) are prevalent in binary classification tasks, they are less applicable to the multi-label classification task of identifying vulnerable libraries.


\section{Related Work}
\subsection{Vulnerable Library Identification}
In the task of identifying vulnerable libraries, there are two main categories of approaches~\cite{viem, anwar2021cleaning, jo2022vulcan, kuehn2021ovana, fastxml, lightxml, chronos, wu2023understanding}, NER-based and XML-based approaches.
VIEM~\cite{viem} is a representative NER-based approach for this task.
It collects the textual description from a vulnerability report and then uses a lightweight NER model~\cite{lample2016neural, yang2017transfer} to identify library entities from this description.
Then it takes a dictionary of library names to match the library entities with this library-name dictionary.
FastXML~\cite{fastxml, prabhu2014fastxml} is a representative XML-based approach for this task.
FastXML uses trees to represent hierarchies over the feature space.
It recursively partitions the parent node by optimizing the normalized Discounted Cumulative Gain (NDCG) as its ranking loss function and returns the ranked list of the most frequently occurring labels in all the leaf nodes for prediction. 
LightXML~\cite{lightxml, jiang2021lightxml}, is a recent deep-learning XML-based approach for this task.
It fine-tunes transformer-based models with dynamic negative label sampling. 
It consists of four components: label clustering, text representation, label recalling, and label ranking.
However, these approaches suffer from high negative results and cannot identify zero-shot libraries due to considering only the names of vulnerable libraries.
On the contrary, \detector{} avoids negative results and identifies zero-shot libraries by correlating library descriptions to vulnerability descriptions.


\subsection{Vulnerable Library Version Identification}
Identifying the vulnerable version of libraries is another important task to help developers avoid vulnerabilities.
These approaches~\cite{Atvhunter, Libid, libscout, libpecker, libdb} need to analyze the source code of different versions of a given library for the natural-language descriptions are not accurate enough to describe the difference between them.
ATVHunter~\cite{Atvhunter}, is a representative approach for identifying the affected version of Android libraries.
ATVHunter extracts the control-flow graph (CFG) as the coarse-grained feature of libraries and conducts a fuzzy hashing technique to extract the features of library methods as the fine-grained feature.
Thus, it combines these features to search its database to identify vulnerable versions.
After identifying the vulnerable libraries, these approaches can help software developers avoid vulnerabilities more accurately.


\subsection{Patch Identification}
Patch identification~\cite{vulfixminer, vulcurator, espi, zhou2021spi, xu2021tracer} identifies whether a code commit corresponds to a vulnerability fix.
It helps users to be aware of vulnerability fixes and apply fixes in time because a vulnerability in open source software (OSS) is suggested to be fixed ``silently'' until the vulnerability is disclosed. 
VulFixMiner~\cite{vulfixminer}, is a representative approach identifying Java and Python vulnerability fixes.
Given a code commit, VulFixMiner extracts the removed and added codes, and uses a pre-trained model, CodeBert~\cite{feng2020codebert}, to encode them.
Then, it uses a fully connected layer to determine whether this code commit corresponds to a vulnerability fix.
These approaches can help identify vulnerable libraries by complementing the description of these 'silently' fixed vulnerabilities.
For example, VulFixMiner has mined 29 commits confirmed by security experts as vulnerability fixes. However, their results require senior experts to determine their corresponding vulnerabilities, thus inducing high costs.

\section{conclusion}\label{sec:conclusion}
In this paper, we have presented our work, being the first to identify vulnerable libraries from descriptions of both vulnerabilities and libraries.
We have designed a TF-IDF matcher to efficiently screen out a set of candidate libraries and a BERT-FNN model to effectively identify the affected libraries for a given vulnerability.
We have constructed a new dataset that is collected from open-source databases and manually validated by senior software engineers, including 5,028 Java $\langle CVE, Library \rangle$ pairs.
We have conducted a comprehensive evaluation of demonstrating \detector{}'s effectiveness and efficiency for library identification, achieving an average F1 score of 0.657, while the state-of-the-art/practice approaches achieve only 0.521.




\bibliographystyle{ACM-Reference-Format}
\bibliography{sample-base}


\begin{thebibliography}{56}


\ifx \showCODEN    \undefined \def \showCODEN     #1{\unskip}     \fi
\ifx \showDOI      \undefined \def \showDOI       #1{#1}\fi
\ifx \showISBNx    \undefined \def \showISBNx     #1{\unskip}     \fi
\ifx \showISBNxiii \undefined \def \showISBNxiii  #1{\unskip}     \fi
\ifx \showISSN     \undefined \def \showISSN      #1{\unskip}     \fi
\ifx \showLCCN     \undefined \def \showLCCN      #1{\unskip}     \fi
\ifx \shownote     \undefined \def \shownote      #1{#1}          \fi
\ifx \showarticletitle \undefined \def \showarticletitle #1{#1}   \fi
\ifx \showURL      \undefined \def \showURL       {\relax}        \fi
\providecommand\bibfield[2]{#2}
\providecommand\bibinfo[2]{#2}
\providecommand\natexlab[1]{#1}
\providecommand\showeprint[2][]{arXiv:#2}

\bibitem[\protect\citeauthoryear{1353}{1353}{2019}]%
        {security1353}
\bibfield{author}{\bibinfo{person}{SECURITY 1353}.} \bibinfo{year}{2019}\natexlab{}.
\newblock \showarticletitle{https://www.jenkins.io/security/advisory/2019-03-25/\#SECURITY-1353}.
\newblock


\bibitem[\protect\citeauthoryear{Advisory}{Advisory}{2022}]%
        {githubAD}
\bibfield{author}{\bibinfo{person}{GitHub Advisory}.} \bibinfo{year}{2022}\natexlab{}.
\newblock \showarticletitle{https://github.com/advisories}.
\newblock


\bibitem[\protect\citeauthoryear{Anwar, Abusnaina, Chen, Li, and Mohaisen}{Anwar et~al\mbox{.}}{2021}]%
        {anwar2021cleaning}
\bibfield{author}{\bibinfo{person}{Afsah Anwar}, \bibinfo{person}{Ahmed Abusnaina}, \bibinfo{person}{Songqing Chen}, \bibinfo{person}{Frank Li}, {and} \bibinfo{person}{David Mohaisen}.} \bibinfo{year}{2021}\natexlab{}.
\newblock \showarticletitle{Cleaning the NVD: Comprehensive quality assessment, improvements, and analyses}.
\newblock \bibinfo{journal}{\emph{IEEE Transactions on Dependable and Secure Computing}} \bibinfo{volume}{19}, \bibinfo{number}{6} (\bibinfo{year}{2021}), \bibinfo{pages}{4255--4269}.
\newblock


\bibitem[\protect\citeauthoryear{Backes, Bugiel, and Derr}{Backes et~al\mbox{.}}{2016}]%
        {libscout}
\bibfield{author}{\bibinfo{person}{Michael Backes}, \bibinfo{person}{Sven Bugiel}, {and} \bibinfo{person}{Erik Derr}.} \bibinfo{year}{2016}\natexlab{}.
\newblock \showarticletitle{Reliable third-party library detection in android and its security applications}. In \bibinfo{booktitle}{\emph{Proceedings of the 2016 ACM SIGSAC Conference on Computer and Communications Security}}. \bibinfo{pages}{356--367}.
\newblock


\bibitem[\protect\citeauthoryear{ChatGPT}{ChatGPT}{2023}]%
        {chatgpt}
\bibfield{author}{\bibinfo{person}{ChatGPT}.} \bibinfo{year}{2023}\natexlab{}.
\newblock \showarticletitle{https://chat.openai.com/}.
\newblock


\bibitem[\protect\citeauthoryear{Chen, Santosa, Sharma, and Lo}{Chen et~al\mbox{.}}{2020a}]%
        {fastxml}
\bibfield{author}{\bibinfo{person}{Yang Chen}, \bibinfo{person}{Andrew~E Santosa}, \bibinfo{person}{Asankhaya Sharma}, {and} \bibinfo{person}{David Lo}.} \bibinfo{year}{2020}\natexlab{a}.
\newblock \showarticletitle{Automated identification of libraries from vulnerability data}. In \bibinfo{booktitle}{\emph{Proceedings of the ACM/IEEE 42nd International Conference on Software Engineering: Software Engineering in Practice}}. \bibinfo{pages}{90--99}.
\newblock


\bibitem[\protect\citeauthoryear{Chen, Santosa, Yi, Sharma, Sharma, and Lo}{Chen et~al\mbox{.}}{2020b}]%
        {chen2020machine}
\bibfield{author}{\bibinfo{person}{Yang Chen}, \bibinfo{person}{Andrew~E Santosa}, \bibinfo{person}{Ang~Ming Yi}, \bibinfo{person}{Abhishek Sharma}, \bibinfo{person}{Asankhaya Sharma}, {and} \bibinfo{person}{David Lo}.} \bibinfo{year}{2020}\natexlab{b}.
\newblock \showarticletitle{A machine learning approach for vulnerability curation}. In \bibinfo{booktitle}{\emph{Proceedings of the 17th International Conference on Mining Software Repositories}}. \bibinfo{pages}{32--42}.
\newblock


\bibitem[\protect\citeauthoryear{CPE}{CPE}{2022}]%
        {cpe}
\bibfield{author}{\bibinfo{person}{CPE}.} \bibinfo{year}{2022}\natexlab{}.
\newblock \showarticletitle{https://nvd.nist.gov/products/cpe}.
\newblock


\bibitem[\protect\citeauthoryear{De~Boer, Kroese, Mannor, and Rubinstein}{De~Boer et~al\mbox{.}}{2005}]%
        {de2005tutorial}
\bibfield{author}{\bibinfo{person}{Pieter-Tjerk De~Boer}, \bibinfo{person}{Dirk~P Kroese}, \bibinfo{person}{Shie Mannor}, {and} \bibinfo{person}{Reuven~Y Rubinstein}.} \bibinfo{year}{2005}\natexlab{}.
\newblock \showarticletitle{A tutorial on the cross-entropy method}.
\newblock \bibinfo{journal}{\emph{Annals of operations research}} \bibinfo{volume}{134}, \bibinfo{number}{1} (\bibinfo{year}{2005}), \bibinfo{pages}{19--67}.
\newblock


\bibitem[\protect\citeauthoryear{Devlin, Chang, Lee, and Toutanova}{Devlin et~al\mbox{.}}{2018}]%
        {bert}
\bibfield{author}{\bibinfo{person}{Jacob Devlin}, \bibinfo{person}{Ming-Wei Chang}, \bibinfo{person}{Kenton Lee}, {and} \bibinfo{person}{Kristina Toutanova}.} \bibinfo{year}{2018}\natexlab{}.
\newblock \showarticletitle{Bert: Pre-training of deep bidirectional transformers for language understanding}.
\newblock \bibinfo{journal}{\emph{arXiv preprint arXiv:1810.04805}} (\bibinfo{year}{2018}).
\newblock


\bibitem[\protect\citeauthoryear{Dong, Guo, Chen, Xing, Zhang, and Wang}{Dong et~al\mbox{.}}{2019}]%
        {viem}
\bibfield{author}{\bibinfo{person}{Ying Dong}, \bibinfo{person}{Wenbo Guo}, \bibinfo{person}{Yueqi Chen}, \bibinfo{person}{Xinyu Xing}, \bibinfo{person}{Yuqing Zhang}, {and} \bibinfo{person}{Gang Wang}.} \bibinfo{year}{2019}\natexlab{}.
\newblock \showarticletitle{Towards the detection of inconsistencies in public security vulnerability reports}. In \bibinfo{booktitle}{\emph{28th USENIX security symposium (USENIX Security 19)}}. \bibinfo{pages}{869--885}.
\newblock


\bibitem[\protect\citeauthoryear{Feng, Guo, Tang, Duan, Feng, Gong, Shou, Qin, Liu, Jiang, et~al\mbox{.}}{Feng et~al\mbox{.}}{2020}]%
        {feng2020codebert}
\bibfield{author}{\bibinfo{person}{Zhangyin Feng}, \bibinfo{person}{Daya Guo}, \bibinfo{person}{Duyu Tang}, \bibinfo{person}{Nan Duan}, \bibinfo{person}{Xiaocheng Feng}, \bibinfo{person}{Ming Gong}, \bibinfo{person}{Linjun Shou}, \bibinfo{person}{Bing Qin}, \bibinfo{person}{Ting Liu}, \bibinfo{person}{Daxin Jiang}, {et~al\mbox{.}}} \bibinfo{year}{2020}\natexlab{}.
\newblock \showarticletitle{Codebert: A pre-trained model for programming and natural languages}.
\newblock \bibinfo{journal}{\emph{arXiv preprint arXiv:2002.08155}} (\bibinfo{year}{2020}).
\newblock


\bibitem[\protect\citeauthoryear{Finkel, Grenager, and Manning}{Finkel et~al\mbox{.}}{2005}]%
        {finkel2005incorporating}
\bibfield{author}{\bibinfo{person}{Jenny~Rose Finkel}, \bibinfo{person}{Trond Grenager}, {and} \bibinfo{person}{Christopher~D Manning}.} \bibinfo{year}{2005}\natexlab{}.
\newblock \showarticletitle{Incorporating non-local information into information extraction systems by gibbs sampling}. In \bibinfo{booktitle}{\emph{Proceedings of the 43rd annual meeting of the association for computational linguistics (ACL’05)}}. \bibinfo{pages}{363--370}.
\newblock


\bibitem[\protect\citeauthoryear{Ghag and Shah}{Ghag and Shah}{2015}]%
        {ghag2015comparative}
\bibfield{author}{\bibinfo{person}{Kranti~Vithal Ghag} {and} \bibinfo{person}{Ketan Shah}.} \bibinfo{year}{2015}\natexlab{}.
\newblock \showarticletitle{Comparative analysis of effect of stopwords removal on sentiment classification}. In \bibinfo{booktitle}{\emph{2015 international conference on computer, communication and control (IC4)}}. IEEE, \bibinfo{pages}{1--6}.
\newblock


\bibitem[\protect\citeauthoryear{Gupta, Bohra, Prabhu, Purohit, and Varma}{Gupta et~al\mbox{.}}{2021}]%
        {zestxml}
\bibfield{author}{\bibinfo{person}{Nilesh Gupta}, \bibinfo{person}{Sakina Bohra}, \bibinfo{person}{Yashoteja Prabhu}, \bibinfo{person}{Saurabh Purohit}, {and} \bibinfo{person}{Manik Varma}.} \bibinfo{year}{2021}\natexlab{}.
\newblock \showarticletitle{Generalized zero-shot extreme multi-label learning}. In \bibinfo{booktitle}{\emph{Proceedings of the 27th ACM SIGKDD Conference on Knowledge Discovery \& Data Mining}}. \bibinfo{pages}{527--535}.
\newblock


\bibitem[\protect\citeauthoryear{Han and Moraga}{Han and Moraga}{1995}]%
        {sigmoid}
\bibfield{author}{\bibinfo{person}{Jun Han} {and} \bibinfo{person}{Claudio Moraga}.} \bibinfo{year}{1995}\natexlab{}.
\newblock \showarticletitle{The influence of the sigmoid function parameters on the speed of backpropagation learning}. In \bibinfo{booktitle}{\emph{International workshop on artificial neural networks}}. Springer, \bibinfo{pages}{195--201}.
\newblock


\bibitem[\protect\citeauthoryear{Haryono, Kang, Sharma, Sharma, Santosa, Yi, and Lo}{Haryono et~al\mbox{.}}{2022}]%
        {lightxml}
\bibfield{author}{\bibinfo{person}{Stefanus~A Haryono}, \bibinfo{person}{Hong~Jin Kang}, \bibinfo{person}{Abhishek Sharma}, \bibinfo{person}{Asankhaya Sharma}, \bibinfo{person}{Andrew Santosa}, \bibinfo{person}{Ang~Ming Yi}, {and} \bibinfo{person}{David Lo}.} \bibinfo{year}{2022}\natexlab{}.
\newblock \showarticletitle{Automated Identification of Libraries from Vulnerability Data: Can We Do Better?}
\newblock  (\bibinfo{year}{2022}).
\newblock


\bibitem[\protect\citeauthoryear{Jiang, Wang, Sun, Yang, Zhao, and Zhuang}{Jiang et~al\mbox{.}}{2021}]%
        {jiang2021lightxml}
\bibfield{author}{\bibinfo{person}{Ting Jiang}, \bibinfo{person}{Deqing Wang}, \bibinfo{person}{Leilei Sun}, \bibinfo{person}{Huayi Yang}, \bibinfo{person}{Zhengyang Zhao}, {and} \bibinfo{person}{Fuzhen Zhuang}.} \bibinfo{year}{2021}\natexlab{}.
\newblock \showarticletitle{Lightxml: Transformer with dynamic negative sampling for high-performance extreme multi-label text classification}. In \bibinfo{booktitle}{\emph{Proceedings of the AAAI Conference on Artificial Intelligence}}, Vol.~\bibinfo{volume}{35}. \bibinfo{pages}{7987--7994}.
\newblock


\bibitem[\protect\citeauthoryear{Jo, Lee, and Shin}{Jo et~al\mbox{.}}{2022}]%
        {jo2022vulcan}
\bibfield{author}{\bibinfo{person}{Hyeonseong Jo}, \bibinfo{person}{Yongjae Lee}, {and} \bibinfo{person}{Seungwon Shin}.} \bibinfo{year}{2022}\natexlab{}.
\newblock \showarticletitle{Vulcan: Automatic extraction and analysis of cyber threat intelligence from unstructured text}.
\newblock \bibinfo{journal}{\emph{Computers \& Security}}  \bibinfo{volume}{120} (\bibinfo{year}{2022}), \bibinfo{pages}{102763}.
\newblock


\bibitem[\protect\citeauthoryear{Jones}{Jones}{1972}]%
        {jones1972statistical}
\bibfield{author}{\bibinfo{person}{Karen~Sparck Jones}.} \bibinfo{year}{1972}\natexlab{}.
\newblock \showarticletitle{A statistical interpretation of term specificity and its application in retrieval}.
\newblock \bibinfo{journal}{\emph{Journal of documentation}} (\bibinfo{year}{1972}).
\newblock


\bibitem[\protect\citeauthoryear{Jones}{Jones}{2004}]%
        {jones2004IDF}
\bibfield{author}{\bibinfo{person}{Karen~Sp{\"a}rck Jones}.} \bibinfo{year}{2004}\natexlab{}.
\newblock \showarticletitle{IDF term weighting and IR research lessons}.
\newblock \bibinfo{journal}{\emph{Journal of documentation}} (\bibinfo{year}{2004}).
\newblock


\bibitem[\protect\citeauthoryear{Kasauli, Knauss, Horkoff, Liebel, and de~Oliveira~Neto}{Kasauli et~al\mbox{.}}{2021}]%
        {kasauli2021requirements}
\bibfield{author}{\bibinfo{person}{Rashidah Kasauli}, \bibinfo{person}{Eric Knauss}, \bibinfo{person}{Jennifer Horkoff}, \bibinfo{person}{Grischa Liebel}, {and} \bibinfo{person}{Francisco~Gomes de Oliveira~Neto}.} \bibinfo{year}{2021}\natexlab{}.
\newblock \showarticletitle{Requirements engineering challenges and practices in large-scale agile system development}.
\newblock \bibinfo{journal}{\emph{Journal of Systems and Software}}  \bibinfo{volume}{172} (\bibinfo{year}{2021}), \bibinfo{pages}{110851}.
\newblock


\bibitem[\protect\citeauthoryear{Kaur and Buttar}{Kaur and Buttar}{2018}]%
        {kaur2018systematic}
\bibfield{author}{\bibinfo{person}{Jashanjot Kaur} {and} \bibinfo{person}{P~Kaur Buttar}.} \bibinfo{year}{2018}\natexlab{}.
\newblock \showarticletitle{A systematic review on stopword removal algorithms}.
\newblock \bibinfo{journal}{\emph{International Journal on Future Revolution in Computer Science \& Communication Engineering}} \bibinfo{volume}{4}, \bibinfo{number}{4} (\bibinfo{year}{2018}), \bibinfo{pages}{207--210}.
\newblock


\bibitem[\protect\citeauthoryear{Khandagale, Xiao, and Babbar}{Khandagale et~al\mbox{.}}{2020}]%
        {khandagale2020bonsai}
\bibfield{author}{\bibinfo{person}{Sujay Khandagale}, \bibinfo{person}{Han Xiao}, {and} \bibinfo{person}{Rohit Babbar}.} \bibinfo{year}{2020}\natexlab{}.
\newblock \showarticletitle{Bonsai: diverse and shallow trees for extreme multi-label classification}.
\newblock \bibinfo{journal}{\emph{Machine Learning}} \bibinfo{volume}{109}, \bibinfo{number}{11} (\bibinfo{year}{2020}), \bibinfo{pages}{2099--2119}.
\newblock


\bibitem[\protect\citeauthoryear{Kuehn, Bayer, Wendelborn, and Reuter}{Kuehn et~al\mbox{.}}{2021}]%
        {kuehn2021ovana}
\bibfield{author}{\bibinfo{person}{Philipp Kuehn}, \bibinfo{person}{Markus Bayer}, \bibinfo{person}{Marc Wendelborn}, {and} \bibinfo{person}{Christian Reuter}.} \bibinfo{year}{2021}\natexlab{}.
\newblock \showarticletitle{OVANA: An approach to analyze and improve the information quality of vulnerability databases}. In \bibinfo{booktitle}{\emph{Proceedings of the 16th International Conference on Availability, Reliability and Security}}. \bibinfo{pages}{1--11}.
\newblock


\bibitem[\protect\citeauthoryear{Lample, Ballesteros, Subramanian, Kawakami, and Dyer}{Lample et~al\mbox{.}}{2016}]%
        {lample2016neural}
\bibfield{author}{\bibinfo{person}{Guillaume Lample}, \bibinfo{person}{Miguel Ballesteros}, \bibinfo{person}{Sandeep Subramanian}, \bibinfo{person}{Kazuya Kawakami}, {and} \bibinfo{person}{Chris Dyer}.} \bibinfo{year}{2016}\natexlab{}.
\newblock \showarticletitle{Neural architectures for named entity recognition}.
\newblock \bibinfo{journal}{\emph{arXiv preprint arXiv:1603.01360}} (\bibinfo{year}{2016}).
\newblock


\bibitem[\protect\citeauthoryear{Li, Allal, Zi, Muennighoff, Kocetkov, Mou, Marone, Akiki, Li, Chim, et~al\mbox{.}}{Li et~al\mbox{.}}{2023}]%
        {starcoder}
\bibfield{author}{\bibinfo{person}{Raymond Li}, \bibinfo{person}{Loubna~Ben Allal}, \bibinfo{person}{Yangtian Zi}, \bibinfo{person}{Niklas Muennighoff}, \bibinfo{person}{Denis Kocetkov}, \bibinfo{person}{Chenghao Mou}, \bibinfo{person}{Marc Marone}, \bibinfo{person}{Christopher Akiki}, \bibinfo{person}{Jia Li}, \bibinfo{person}{Jenny Chim}, {et~al\mbox{.}}} \bibinfo{year}{2023}\natexlab{}.
\newblock \showarticletitle{StarCoder: may the source be with you!}
\newblock \bibinfo{journal}{\emph{arXiv preprint arXiv:2305.06161}} (\bibinfo{year}{2023}).
\newblock


\bibitem[\protect\citeauthoryear{Lyu, Le-Cong, Kang, Widyasari, Zhao, Le, Li, and Lo}{Lyu et~al\mbox{.}}{2023}]%
        {chronos}
\bibfield{author}{\bibinfo{person}{Yunbo Lyu}, \bibinfo{person}{Thanh Le-Cong}, \bibinfo{person}{Hong~Jin Kang}, \bibinfo{person}{Ratnadira Widyasari}, \bibinfo{person}{Zhipeng Zhao}, \bibinfo{person}{Xuan-Bach~D Le}, \bibinfo{person}{Ming Li}, {and} \bibinfo{person}{David Lo}.} \bibinfo{year}{2023}\natexlab{}.
\newblock \showarticletitle{Chronos: Time-aware zero-shot identification of libraries from vulnerability reports}.
\newblock \bibinfo{journal}{\emph{arXiv preprint arXiv:2301.03944}} (\bibinfo{year}{2023}).
\newblock


\bibitem[\protect\citeauthoryear{Maven}{Maven}{2022}]%
        {maven}
\bibfield{author}{\bibinfo{person}{Maven}.} \bibinfo{year}{2022}\natexlab{}.
\newblock \showarticletitle{https://maven.apache.org}.
\newblock


\bibitem[\protect\citeauthoryear{Meng, Nagy, Yao, Zhuang, and Argoty}{Meng et~al\mbox{.}}{2018}]%
        {meng2018secure}
\bibfield{author}{\bibinfo{person}{Na Meng}, \bibinfo{person}{Stefan Nagy}, \bibinfo{person}{Danfeng Yao}, \bibinfo{person}{Wenjie Zhuang}, {and} \bibinfo{person}{Gustavo~Arango Argoty}.} \bibinfo{year}{2018}\natexlab{}.
\newblock \showarticletitle{Secure coding practices in java: Challenges and vulnerabilities}. In \bibinfo{booktitle}{\emph{Proceedings of the 40th International Conference on Software Engineering}}. \bibinfo{pages}{372--383}.
\newblock


\bibitem[\protect\citeauthoryear{Mikolov, Sutskever, Chen, Corrado, and Dean}{Mikolov et~al\mbox{.}}{2013}]%
        {mikolov2013distributed}
\bibfield{author}{\bibinfo{person}{Tomas Mikolov}, \bibinfo{person}{Ilya Sutskever}, \bibinfo{person}{Kai Chen}, \bibinfo{person}{Greg~S Corrado}, {and} \bibinfo{person}{Jeff Dean}.} \bibinfo{year}{2013}\natexlab{}.
\newblock \showarticletitle{Distributed representations of words and phrases and their compositionality}.
\newblock \bibinfo{journal}{\emph{Advances in neural information processing systems}}  \bibinfo{volume}{26} (\bibinfo{year}{2013}).
\newblock


\bibitem[\protect\citeauthoryear{Narayan, Gupta, Khan, Khan, Shao, and Shah}{Narayan et~al\mbox{.}}{2021}]%
        {narayan2021discriminative}
\bibfield{author}{\bibinfo{person}{Sanath Narayan}, \bibinfo{person}{Akshita Gupta}, \bibinfo{person}{Salman Khan}, \bibinfo{person}{Fahad~Shahbaz Khan}, \bibinfo{person}{Ling Shao}, {and} \bibinfo{person}{Mubarak Shah}.} \bibinfo{year}{2021}\natexlab{}.
\newblock \showarticletitle{Discriminative region-based multi-label zero-shot learning}. In \bibinfo{booktitle}{\emph{Proceedings of the IEEE/CVF international conference on computer vision}}. \bibinfo{pages}{8731--8740}.
\newblock


\bibitem[\protect\citeauthoryear{Nguyen, Le-Cong, Kang, Le, and Lo}{Nguyen et~al\mbox{.}}{2022}]%
        {vulcurator}
\bibfield{author}{\bibinfo{person}{Truong~Giang Nguyen}, \bibinfo{person}{Thanh Le-Cong}, \bibinfo{person}{Hong~Jin Kang}, \bibinfo{person}{Xuan-Bach~D Le}, {and} \bibinfo{person}{David Lo}.} \bibinfo{year}{2022}\natexlab{}.
\newblock \showarticletitle{VulCurator: a vulnerability-fixing commit detector}. In \bibinfo{booktitle}{\emph{Proceedings of the 30th ACM Joint European Software Engineering Conference and Symposium on the Foundations of Software Engineering}}. \bibinfo{pages}{1726--1730}.
\newblock


\bibitem[\protect\citeauthoryear{NVD}{NVD}{2022}]%
        {nvd}
\bibfield{author}{\bibinfo{person}{NVD}.} \bibinfo{year}{2022}\natexlab{}.
\newblock \showarticletitle{https://nvd.nist.gov}.
\newblock


\bibitem[\protect\citeauthoryear{Pham, Nguyen, Nguyen, and Nguyen}{Pham et~al\mbox{.}}{2010}]%
        {pham2010detection}
\bibfield{author}{\bibinfo{person}{Nam~H Pham}, \bibinfo{person}{Tung~Thanh Nguyen}, \bibinfo{person}{Hoan~Anh Nguyen}, {and} \bibinfo{person}{Tien~N Nguyen}.} \bibinfo{year}{2010}\natexlab{}.
\newblock \showarticletitle{Detection of recurring software vulnerabilities}. In \bibinfo{booktitle}{\emph{Proceedings of the IEEE/ACM international conference on Automated software engineering}}. \bibinfo{pages}{447--456}.
\newblock


\bibitem[\protect\citeauthoryear{Prabhu and Varma}{Prabhu and Varma}{2014}]%
        {prabhu2014fastxml}
\bibfield{author}{\bibinfo{person}{Yashoteja Prabhu} {and} \bibinfo{person}{Manik Varma}.} \bibinfo{year}{2014}\natexlab{}.
\newblock \showarticletitle{Fastxml: A fast, accurate and stable tree-classifier for extreme multi-label learning}. In \bibinfo{booktitle}{\emph{Proceedings of the 20th ACM SIGKDD international conference on Knowledge discovery and data mining}}. \bibinfo{pages}{263--272}.
\newblock


\bibitem[\protect\citeauthoryear{Prana, Sharma, Shar, Foo, Santosa, Sharma, and Lo}{Prana et~al\mbox{.}}{2021}]%
        {prana2021out}
\bibfield{author}{\bibinfo{person}{Gede Artha~Azriadi Prana}, \bibinfo{person}{Abhishek Sharma}, \bibinfo{person}{Lwin~Khin Shar}, \bibinfo{person}{Darius Foo}, \bibinfo{person}{Andrew~E Santosa}, \bibinfo{person}{Asankhaya Sharma}, {and} \bibinfo{person}{David Lo}.} \bibinfo{year}{2021}\natexlab{}.
\newblock \showarticletitle{Out of sight, out of mind? How vulnerable dependencies affect open-source projects}.
\newblock \bibinfo{journal}{\emph{Empirical Software Engineering}} \bibinfo{volume}{26}, \bibinfo{number}{4} (\bibinfo{year}{2021}), \bibinfo{pages}{1--34}.
\newblock


\bibitem[\protect\citeauthoryear{PyPI}{PyPI}{2023}]%
        {pypi}
\bibfield{author}{\bibinfo{person}{PyPI}.} \bibinfo{year}{2023}\natexlab{}.
\newblock \showarticletitle{https://pypi.org/}.
\newblock


\bibitem[\protect\citeauthoryear{Report}{Report}{2022}]%
        {nvd2022report}
\bibfield{author}{\bibinfo{person}{NVD Report}.} \bibinfo{year}{2022}\natexlab{}.
\newblock \showarticletitle{https://www.reversinglabs.com/blog/nvd-analysis-2022-why-you-need-to-modernize-your-software-security-approach}.
\newblock


\bibitem[\protect\citeauthoryear{Schmid}{Schmid}{1994}]%
        {schmid1994part}
\bibfield{author}{\bibinfo{person}{Helmut Schmid}.} \bibinfo{year}{1994}\natexlab{}.
\newblock \showarticletitle{Part-of-speech tagging with neural networks}.
\newblock \bibinfo{journal}{\emph{arXiv preprint cmp-lg/9410018}} (\bibinfo{year}{1994}).
\newblock


\bibitem[\protect\citeauthoryear{Shen, Wang, and Han}{Shen et~al\mbox{.}}{2014}]%
        {shen2014entity}
\bibfield{author}{\bibinfo{person}{Wei Shen}, \bibinfo{person}{Jianyong Wang}, {and} \bibinfo{person}{Jiawei Han}.} \bibinfo{year}{2014}\natexlab{}.
\newblock \showarticletitle{Entity linking with a knowledge base: Issues, techniques, and solutions}.
\newblock \bibinfo{journal}{\emph{IEEE Transactions on Knowledge and Data Engineering}} \bibinfo{volume}{27}, \bibinfo{number}{2} (\bibinfo{year}{2014}), \bibinfo{pages}{443--460}.
\newblock


\bibitem[\protect\citeauthoryear{Tang, Wang, Zhang, Han, Luo, and Zhang}{Tang et~al\mbox{.}}{2022}]%
        {libdb}
\bibfield{author}{\bibinfo{person}{Wei Tang}, \bibinfo{person}{Yanlin Wang}, \bibinfo{person}{Hongyu Zhang}, \bibinfo{person}{Shi Han}, \bibinfo{person}{Ping Luo}, {and} \bibinfo{person}{Dongmei Zhang}.} \bibinfo{year}{2022}\natexlab{}.
\newblock \showarticletitle{LibDB: An Effective and Efficient Framework for Detecting Third-Party Libraries in Binaries}.
\newblock \bibinfo{journal}{\emph{arXiv preprint arXiv:2204.10232}} (\bibinfo{year}{2022}).
\newblock


\bibitem[\protect\citeauthoryear{Wang, Chen, Huang, Shi, Xu, Peng, Wu, and Liu}{Wang et~al\mbox{.}}{2020}]%
        {wang2020empirical}
\bibfield{author}{\bibinfo{person}{Ying Wang}, \bibinfo{person}{Bihuan Chen}, \bibinfo{person}{Kaifeng Huang}, \bibinfo{person}{Bowen Shi}, \bibinfo{person}{Congying Xu}, \bibinfo{person}{Xin Peng}, \bibinfo{person}{Yijian Wu}, {and} \bibinfo{person}{Yang Liu}.} \bibinfo{year}{2020}\natexlab{}.
\newblock \showarticletitle{An empirical study of usages, updates and risks of third-party libraries in java projects}. In \bibinfo{booktitle}{\emph{2020 IEEE International Conference on Software Maintenance and Evolution (ICSME)}}. IEEE, \bibinfo{pages}{35--45}.
\newblock


\bibitem[\protect\citeauthoryear{Wu, Liu, Feng, Xie, Siow, and Lin}{Wu et~al\mbox{.}}{2022}]%
        {espi}
\bibfield{author}{\bibinfo{person}{Bozhi Wu}, \bibinfo{person}{Shangqing Liu}, \bibinfo{person}{Ruitao Feng}, \bibinfo{person}{Xiaofei Xie}, \bibinfo{person}{Jingkai Siow}, {and} \bibinfo{person}{Shang-Wei Lin}.} \bibinfo{year}{2022}\natexlab{}.
\newblock \showarticletitle{Enhancing Security Patch Identification by Capturing Structures in Commits}.
\newblock \bibinfo{journal}{\emph{IEEE Transactions on Dependable and Secure Computing}} (\bibinfo{year}{2022}).
\newblock


\bibitem[\protect\citeauthoryear{Wu, Jiang, Jiang, Xie, and Tu}{Wu et~al\mbox{.}}{2023a}]%
        {wu2023plms}
\bibfield{author}{\bibinfo{person}{Weiqi Wu}, \bibinfo{person}{Chengyue Jiang}, \bibinfo{person}{Yong Jiang}, \bibinfo{person}{Pengjun Xie}, {and} \bibinfo{person}{Kewei Tu}.} \bibinfo{year}{2023}\natexlab{a}.
\newblock \showarticletitle{Do PLMs Know and Understand Ontological Knowledge?}. In \bibinfo{booktitle}{\emph{Proceedings of the 61st Annual Meeting of the Association for Computational Linguistics (Volume 1: Long Papers)}}. \bibinfo{pages}{3080--3101}.
\newblock


\bibitem[\protect\citeauthoryear{Wu, Yu, Wen, Li, Zou, and Jin}{Wu et~al\mbox{.}}{2023b}]%
        {wu2023understanding}
\bibfield{author}{\bibinfo{person}{Yulun Wu}, \bibinfo{person}{Zeliang Yu}, \bibinfo{person}{Ming Wen}, \bibinfo{person}{Qiang Li}, \bibinfo{person}{Deqing Zou}, {and} \bibinfo{person}{Hai Jin}.} \bibinfo{year}{2023}\natexlab{b}.
\newblock \showarticletitle{Understanding the Threats of Upstream Vulnerabilities to Downstream Projects in the Maven Ecosystem}. In \bibinfo{booktitle}{\emph{2023 IEEE/ACM 45th International Conference on Software Engineering (ICSE)}}. IEEE, \bibinfo{pages}{1046--1058}.
\newblock


\bibitem[\protect\citeauthoryear{Wydmuch, Jasinska, Kuznetsov, Busa-Fekete, and Dembczynski}{Wydmuch et~al\mbox{.}}{2018}]%
        {wydmuch2018no}
\bibfield{author}{\bibinfo{person}{Marek Wydmuch}, \bibinfo{person}{Kalina Jasinska}, \bibinfo{person}{Mikhail Kuznetsov}, \bibinfo{person}{R{\'o}bert Busa-Fekete}, {and} \bibinfo{person}{Krzysztof Dembczynski}.} \bibinfo{year}{2018}\natexlab{}.
\newblock \showarticletitle{A no-regret generalization of hierarchical softmax to extreme multi-label classification}.
\newblock \bibinfo{journal}{\emph{Advances in neural information processing systems}}  \bibinfo{volume}{31} (\bibinfo{year}{2018}).
\newblock


\bibitem[\protect\citeauthoryear{Xu, Chen, Lu, Huang, Peng, and Liu}{Xu et~al\mbox{.}}{2021}]%
        {xu2021tracer}
\bibfield{author}{\bibinfo{person}{Congying Xu}, \bibinfo{person}{Bihuan Chen}, \bibinfo{person}{Chenhao Lu}, \bibinfo{person}{Kaifeng Huang}, \bibinfo{person}{Xin Peng}, {and} \bibinfo{person}{Yang Liu}.} \bibinfo{year}{2021}\natexlab{}.
\newblock \showarticletitle{TRACER: Finding Patches for Open Source Software Vulnerabilities}.
\newblock \bibinfo{journal}{\emph{arXiv preprint arXiv:2112.02240}} (\bibinfo{year}{2021}).
\newblock


\bibitem[\protect\citeauthoryear{Yang, Salakhutdinov, and Cohen}{Yang et~al\mbox{.}}{2017}]%
        {yang2017transfer}
\bibfield{author}{\bibinfo{person}{Zhilin Yang}, \bibinfo{person}{Ruslan Salakhutdinov}, {and} \bibinfo{person}{William~W Cohen}.} \bibinfo{year}{2017}\natexlab{}.
\newblock \showarticletitle{Transfer learning for sequence tagging with hierarchical recurrent networks}.
\newblock \bibinfo{journal}{\emph{arXiv preprint arXiv:1703.06345}} (\bibinfo{year}{2017}).
\newblock


\bibitem[\protect\citeauthoryear{Yao and Shepperd}{Yao and Shepperd}{2020}]%
        {yao2020assessing}
\bibfield{author}{\bibinfo{person}{Jingxiu Yao} {and} \bibinfo{person}{Martin Shepperd}.} \bibinfo{year}{2020}\natexlab{}.
\newblock \showarticletitle{Assessing software defection prediction performance: Why using the Matthews correlation coefficient matters}. In \bibinfo{booktitle}{\emph{Proceedings of the 24th International Conference on Evaluation and Assessment in Software Engineering}}. \bibinfo{pages}{120--129}.
\newblock


\bibitem[\protect\citeauthoryear{Zhan, Fan, Chen, We, Liu, Luo, and Liu}{Zhan et~al\mbox{.}}{2021}]%
        {Atvhunter}
\bibfield{author}{\bibinfo{person}{Xian Zhan}, \bibinfo{person}{Lingling Fan}, \bibinfo{person}{Sen Chen}, \bibinfo{person}{Feng We}, \bibinfo{person}{Tianming Liu}, \bibinfo{person}{Xiapu Luo}, {and} \bibinfo{person}{Yang Liu}.} \bibinfo{year}{2021}\natexlab{}.
\newblock \showarticletitle{Atvhunter: Reliable version detection of third-party libraries for vulnerability identification in android applications}. In \bibinfo{booktitle}{\emph{2021 IEEE/ACM 43rd International Conference on Software Engineering (ICSE)}}. IEEE, \bibinfo{pages}{1695--1707}.
\newblock


\bibitem[\protect\citeauthoryear{Zhang, Beresford, and Kollmann}{Zhang et~al\mbox{.}}{2019}]%
        {Libid}
\bibfield{author}{\bibinfo{person}{Jiexin Zhang}, \bibinfo{person}{Alastair~R Beresford}, {and} \bibinfo{person}{Stephan~A Kollmann}.} \bibinfo{year}{2019}\natexlab{}.
\newblock \showarticletitle{Libid: reliable identification of obfuscated third-party android libraries}. In \bibinfo{booktitle}{\emph{Proceedings of the 28th ACM SIGSOFT International Symposium on Software Testing and Analysis}}. \bibinfo{pages}{55--65}.
\newblock


\bibitem[\protect\citeauthoryear{Zhang, Guo, He, and Hu}{Zhang et~al\mbox{.}}{2023}]%
        {zhang2023bidirectional}
\bibfield{author}{\bibinfo{person}{Xing Zhang}, \bibinfo{person}{Guanchen Guo}, \bibinfo{person}{Xiao He}, {and} \bibinfo{person}{Zhenjiang Hu}.} \bibinfo{year}{2023}\natexlab{}.
\newblock \showarticletitle{Bidirectional Object-Oriented Programming: Towards Programmatic and Direct Manipulation of Objects}.
\newblock \bibinfo{journal}{\emph{Proceedings of the ACM on Programming Languages}} \bibinfo{volume}{7}, \bibinfo{number}{OOPSLA1} (\bibinfo{year}{2023}), \bibinfo{pages}{230--255}.
\newblock


\bibitem[\protect\citeauthoryear{Zhang, Dai, Zhang, Huang, Yang, Yang, and Chen}{Zhang et~al\mbox{.}}{2018}]%
        {libpecker}
\bibfield{author}{\bibinfo{person}{Yuan Zhang}, \bibinfo{person}{Jiarun Dai}, \bibinfo{person}{Xiaohan Zhang}, \bibinfo{person}{Sirong Huang}, \bibinfo{person}{Zhemin Yang}, \bibinfo{person}{Min Yang}, {and} \bibinfo{person}{Hao Chen}.} \bibinfo{year}{2018}\natexlab{}.
\newblock \showarticletitle{Detecting third-party libraries in android applications with high precision and recall}. In \bibinfo{booktitle}{\emph{2018 IEEE 25th International Conference on Software Analysis, Evolution and Reengineering (SANER)}}. IEEE, \bibinfo{pages}{141--152}.
\newblock


\bibitem[\protect\citeauthoryear{Zhou, Pacheco, Wan, Xia, Lo, Wang, and Hassan}{Zhou et~al\mbox{.}}{2021a}]%
        {vulfixminer}
\bibfield{author}{\bibinfo{person}{Jiayuan Zhou}, \bibinfo{person}{Michael Pacheco}, \bibinfo{person}{Zhiyuan Wan}, \bibinfo{person}{Xin Xia}, \bibinfo{person}{David Lo}, \bibinfo{person}{Yuan Wang}, {and} \bibinfo{person}{Ahmed~E Hassan}.} \bibinfo{year}{2021}\natexlab{a}.
\newblock \showarticletitle{Finding A Needle in a Haystack: Automated Mining of Silent Vulnerability Fixes}. In \bibinfo{booktitle}{\emph{2021 36th IEEE/ACM International Conference on Automated Software Engineering (ASE)}}. IEEE, \bibinfo{pages}{705--716}.
\newblock


\bibitem[\protect\citeauthoryear{Zhou, Siow, Wang, Liu, and Liu}{Zhou et~al\mbox{.}}{2021b}]%
        {zhou2021spi}
\bibfield{author}{\bibinfo{person}{Yaqin Zhou}, \bibinfo{person}{Jing~Kai Siow}, \bibinfo{person}{Chenyu Wang}, \bibinfo{person}{Shangqing Liu}, {and} \bibinfo{person}{Yang Liu}.} \bibinfo{year}{2021}\natexlab{b}.
\newblock \showarticletitle{SPI: Automated Identification of Security Patches via Commits}.
\newblock \bibinfo{journal}{\emph{ACM Transactions on Software Engineering and Methodology (TOSEM)}} \bibinfo{volume}{31}, \bibinfo{number}{1} (\bibinfo{year}{2021}), \bibinfo{pages}{1--27}.
\newblock


\end{thebibliography}


\end{document}